\documentclass[%
 reprint,
 amsmath,amssymb,
 aps,
]{revtex4-2}

\usepackage{graphicx}
\usepackage{dcolumn}
\usepackage{bm}
\usepackage{chemformula}
\usepackage{chemfig}
\usepackage[version=3]{mhchem}
\usepackage{multirow}
\usepackage{mwe,tikz}
\usepackage[percent]{overpic}
\usepackage{graphicx}
\usepackage{caption}
\usepackage{subcaption}
\usepackage{siunitx}
\usepackage{natbib}
\usepackage{xr}
\usepackage{amsmath}
\usepackage{mathtools}
\usepackage{amssymb}
\usepackage{amsmath, bm}
\usepackage{cleveref}

\begin{document}

\title{Solvent Quality and Nonbiological Oligomer Folding: Revisiting Conventional Paradigms}

\author{Cedrix J. Dongmo Foumthuim}
 \affiliation{INFN, Sezione di Roma Tor Vergata, Via della Ricerca Scientifica 1, 00133 Roma, Italy.}
\author{Tobia Arcangeli}
\affiliation{Dipartimento di Chimica, Materiali e Ingegneria Chimica "Giulio Natta", Politecnico di Milano, Sede Leonardo Edificio 6, Piazza Leonardo da Vinci 32, I-20133 Milano, Italy.}%
\author{Tatjana \v{S}krbi\'{c}}%
\affiliation{Dipartimento di Scienze Molecolari e Nanosistemi, Università Ca' Foscari Venezia, Via Torino 155, 30172, Venezia, Italy.}%
\author{Achille Giacometti}
\email{achille.giacometti@unive.it}
\affiliation{Dipartimento di Scienze Molecolari e Nanosistemi, Università Ca' Foscari Venezia, Via Torino 155, 30172, Venezia, Italy.}%
\affiliation{European Centre for Living Technology (ECLT)
Ca’ Bottacin, Dorsoduro 3911, Calle Crosera 30123 Venice, Italy. }%

\date{\today}

\begin{abstract}
We report on extensive molecular dynamics atomistic simulations of a \textit{meta}-substituted \textit{poly}-phenylacetylene (pPA) foldamer dispersed in three solvents, water \ce{H2O}, cyclohexane \ce{cC6H12}, and \textit{n}-hexane \ce{nC6H14}, and for three oligomer lengths \textit{12mer}, \textit{16mer} and \textit{20mer}. At room temperature, we find a tendency of the pPA foldamer to collapse into a helical structure in all three solvents but with rather different stability character, stable in water, marginally stable in n-hexane, unstable in cyclohexane. In the case of water, the initial and final number of hydrogen bonds of the foldamer with water molecules is found to be unchanged, with no formation of intrachain hydrogen bonding, thus indicating that hydrogen bonding plays no role in the folding process. In all three solvents, the folding is found to be mainly driven by electrostatics, nearly identical in the three cases, and largely dominant compared to van der Waals interactions that are different in the three cases.
 This scenario is also supported by the analysis of distribution of the bond and dihedral angles and by a direct calculation of the solvation and transfer free energies via thermodynamic integration. The different stability in the case of cyclohexane and n-hexane notwithstanding their rather similar chemical composition can be traced back to the different entropy-enthalpy compensation that is found similar for water and n-hexane, and very different for cyclohexane.
 A comparison with the same properties for \textit{poly}-phenylalanine oligomers underscores the crucial differences between pPA and peptides.
 To highlight how these findings can hardly be interpreted in terms of a simple "good" and "poor" solvent picture, a molecular dynamics study of a bead-spring polymer chain in a Lennard-Jones fluid is also included.
\end{abstract}

\maketitle


\section{Introduction}
\label{sec:introduction}
Although the concept of \textit{good} and \textit{poor} solvent is rather clear to anyone working in polymer physics, a universally accepted definition is  surprisingly still lacking. Yet, there are several ways in which this concept can be conveyed, the simplest being the chemical affinity \cite{Doi1988}. If the monomers have stronger chemical affinity with the solvent than with other monomers, the solvent tends to dissolve the polymer  and is then defined to be a good solvent for the polymer. In the opposite limit, monomers tend to be segregated from the solvent and hence the polymer collapses into a globular form. In this case, the solvent is classified as a poor solvent for the polymer. In the intermediate case, the chemical moieties of the monomers and the solvent are very similar and the solvent is defined to be a $\theta$ solvent.  For instance for polystyrene, benzene is a good solvent, cyclohexane a $\theta$ solvent, water a poor solvent.
 A more quantitative definition is based on the Flory-Huggins theory \cite{Doi1988,DeGennes1979} via the dimensionless interaction parameter $\chi$ that measures the relative strength of the monomer-monomer, solvent-solvent, and monomer-solvent interactions. The case $\chi<0$ corresponds to a prevalent attraction between monomers and solvent particles that promotes a coil conformation for the polymer (good solvent), the opposite limit is achieved for $\chi>0$  with a prevalent solvent-solvent and monomer-monomer attraction that promotes the collapse of the polymer  (poor solvent). Finally, $\chi=0$ corresponds to the case where the two competing attractions balance one another and this is denoted as a $\theta$ solvent. An alternative definition, having the additional advantage of being independent of a specific model, hinges on the second virial coefficient, a measure of the excluded volume \cite{Khokhlov2002,Rubinstein2003}, with a positive (negative) value corresponding to good (poor) solvents.

A different, albeit related, concept is the notion of solvent (and solute) polarities \cite{Griffiths1979}. Solutes tend to dissolve easily in solvents with like polarities. For instance, a non-polar solvent, such as oil, is immiscible in a polar solvent, such as water, because their polarities are different. Hence, water would act as poor solvent for oil solutes and oil would act as poor solvent for water solutes. The polarity of a molecular entity, be it solute or solvent, is found to be proportional to its total dipole moment which, in turn, is related to its dielectric constant, with higher dielectric constants assigned to more polar chemical entities.

A third related aspect associated with the solubility of a solute in a solvent, is the solvation free energy. This is defined as the change in free energy in transferring a solute from  vacuum (gas phase) to a solvent \cite{Leach2001}, 
with a negative (positive) value indicating that the process is thermodynamically favorable (unfavorable). In general the solvation free energy is composed by four different components. The first two are the free energy required to form the solute cavity within the solvent and the van der Waals interactions between the solvent and the solute. These are present for all solvents and all solutes. The third term is the electrostatic contribution that is only present for charged and polar solutes and/or solvents, and the last term is an explicit hydrogen-bonding term operating only in the presence of explicit donors and acceptors both in the solvent and in the solute. 

Taken together, these three concepts -- solvent quality, solvent polarity, and solvation free energy, subsum the idea of the solvophobic/solvophilic interactions. They are clearly related, but their relation is far from being obvious.  A polymer (the solute) formed by chemical moieties that tend to avoid contact with the solvent, collapses into folded structure as driven by solvophobic interactions, and it is then classified as a polymer in a poor solvent. Conversely, a polymer with solvophilic interactions tends to remain swollen as a polymer in a good solvent. When the solvent is water, then the more specific definition of hydrophobic and polar (or hydrophilic) interactions are commonly used. Even in this case, the notion of solvophobic/solvophilic interactions lacks of a universally accepted definition and mostly relies on the classification of the chemical moieties. Here too, it is however possible to provide a quantification of these concepts in terms of the difference between the mean force potential $W(r)$ for two solute molecules at distance $r$ and the two-body potential $\phi(r)$ acting between the same two molecules at the same distance \cite{BenNaim2012}.

A simple phenomenological way to relate the above concepts assumes the existence of a hydrophobic scale, quantified for instance by the value of the dielectric constant \cite{Griffiths1979}, with organic solvents as the most hydrophobic and water as the most hydrophilic/polar. A similar scale can then be implemented for the monomers forming the polymer thus classifying the polymer as mostly hydrophobic or mostly hydrophilic/polar. Hence, we can surmise that hydrophobic solvents act as good solvents for hydrophobic polymers and bad solvents for hydrophilic/polar polymers, the converse being true for hydrophilic/polar solvents, justifying the rule of thumb "like-dissolves-like". The overall result, is a close relation between solvent quality and solvent polarity, as it was already noted in the framework of biopolymers \cite{Dongmo2023}. The aim of the present study is to provide a similar perspective also for synthetic polymers.

Biopolymers, such as proteins, are fundamental pillars supporting the molecular foundation of life. They play a cutting edge function in life’s machinery and what make them so unique is their capability to perform cellular functions. However, to perform their biological functions, proteins should fold in well defined three-dimensional shape, owing to their conformational freedom. This folding process is driven by intramolecular interactions and by the requirement of maximizing the solvent entropy whose combination overwhelms the solute-solvent interactions which may promote expansion.

In search for conformational optimization, synthetic polymers usually collapse into a structureless globular structure, unlike biopolymers. However, in 1997 Nelson and coworkers \cite{Nelson1997} reported on a synthetic aromatic hydrocarbon backbone -- the \textit{all-meta}-phenylacetylene foldamer (pPA) undergoing a coil-to-helix transition similarly to biomolecules. Several differences however were also observed for this non-biological foldamer. First, the complete absence of intramolecular hydrogen bonding which, by contrast, is believed to be one of the main stability factor for the native structure in proteins.  Second, its dependence on the chain length, with the existence of a critical number of monomers below which no folding is observed. Finally, the geometry of the helices that for pPA is found to be very different from their biological counterparts (see below).

In the attempt to unravel the physical basis of helix stabilization in water \ce{H2O} at room temperature, Srikanta Sen \cite{Srikanta2002} performed atomistic molecular dynamics simulations to analyze the structure, dynamics and energetics of a pPA in water for both helical and coil conformations. Although he did not detect any intrachain hydrogen bonds in the folded conformer, the oligomer was found to maintain a dynamical stable helical motif in water \ce{H2O} with the following geometrical features : a pitch of nearly 5.5 residue per helical turn; a rise of about 0.69 \AA~per residue; an inner pore and outer surface of diameters of about 10 \AA~and 19~\AA, respectively. These values are very different from those measured in proteins \cite{Banavar2023}. Computational limitations, however, confined this study to an assessment of the stability of a prescribed (extended or collapsed) structure, and prevented a detailed analysis of the full time trajectory.

The present study builds on this work and extends it in several aspects. At the same time, it clarifies the relation between solvent quality, solvent polarity, and solvation free energy in this framework.
The manuscript is then organized as follows. After presenting all the necessary technical machinery in Section \ref{sec:materials}, we present our findings in Section \ref{sec:results} with several subsections dedicated to all the different aspects, and finally present some take home messages in Section \ref{sec:conclusions}.
\section {Materials and Methods}
\label{sec:materials}
Section \ref{subsec:bead_spring} includes the description of a simple generic polymer model in explicit solvents that will be initially used to illustrate the concept of solvent quality; Sections \ref{subsec:molecular} and \ref{subsec:simulations} describe the atomistic description of the pPA foldamer and the corresponding all atoms simulations, while Section \ref{subsec:potential} deals with the calculation of the potential of mean force; Finally, Section \ref{subsec:solvation} further describes the thermodynamic integration required to obtain the solvation free energies.
\subsection{Bead-spring model and Lennard-Jones solvent}
\label{subsec:bead_spring}
The bead-spring model \cite{Kremer1990} was implemented following standard prescriptions (see e.g. Ref. \cite{Huang2021} and references therein). The polymer consists of $N=128$ consecutive beads of diameter $\sigma$ and mass $m$ connected by bonds via a FENE potential
\begin{eqnarray}
\label{sec1:eq1}
\phi_{\text{FENE}} \left(r \right) &=& -\frac{k}{2} R_0^2 \ln \left[1- \left(\frac{r}{R_0} \right)^2\right] \qquad 0 \le r \le R_0 \nonumber \\
&&
\end{eqnarray}
where $k=30 \epsilon/\sigma^2$, and $R_0=1.5 \sigma$. 
The (soft) repulsion of two consecutive beads is described via a Week-Chandler-Anderson Lennard-Jones shifted potential \cite{Weeks1971}
\begin{eqnarray}
\label{sec1:eq2}
\phi_{\text{WCA}} \left(r \right) &=& 4 \epsilon \left[\left( \frac{r}{\sigma}\right)^{12} - \left( \frac{r}{\sigma}\right)^{6} + \frac{1}{4} \right] \qquad 0 \le r \le 2^{1/6} \sigma \nonumber \\
&&
\end{eqnarray}
which is cut-off at $r=2^{1/6} \sigma$ as indicated.
Finally, the interactions between two non-consecutive beads in the chain interact via a standard Lennard-Jones potential 
\begin{eqnarray}
\label{sec1:eq3}
\phi_{\text{LJ}} \left(r \right) &=& 4 \epsilon \left[\left( \frac{\sigma}{r}\right)^{12} - \left( \frac{\sigma}{r}\right)^{6} - \left( \frac{\sigma}{r_c}\right)^{12} + \left( \frac{\sigma}{r_c}\right)^{6} \right] \nonumber \\
&& 0 \le r \le r_c 
\end{eqnarray}
which is cut-off at $r_c=3 \sigma$. 

The solvent is also modeled by soft beads of diameter $\sigma$, mass $m$ and interacting via the Lennard-Jones potential (\ref{sec1:eq3}), and the solvent-monomer interaction is also described by the same Lennard-Jones potential (\ref{sec1:eq3}) but with characteristic energy $\epsilon_{ms}$ replacing $\epsilon$. Throughout this calculation $\sigma$ and $\epsilon$ represent the unit of length and energy respectively, and $m$ is taken  as the unit of masses.

All simulations are performed within the LAMMPS computational suit \cite{Plimpton1995,Thompson2022} using a Verlet algorithm \cite{Verlet1967} with a time step $\Delta t=0.005 \tau$, in LJ units of time $\tau=\sqrt{m \sigma^2/\epsilon}$, with periodic boundary conditions applied in all directions. Temperature is controlled by a Nos\`e-Hoover thermostat with a damping time $\Gamma=100 \tau$, and pressure is controlled by a Nos\`e-Hoover barostat \cite{Shinoda2004}. Temperatures will be reported in reduced units $T^{*}=k_B T/\epsilon$, $k_B$ being the Boltzmann constant.
\subsection{Atomistic molecular models}
\label{subsec:molecular}

\begin{figure*}[htbp]
\centering
\includegraphics[width=3in, trim=80cm 81.5cm 80cm 81cm, clip=true]{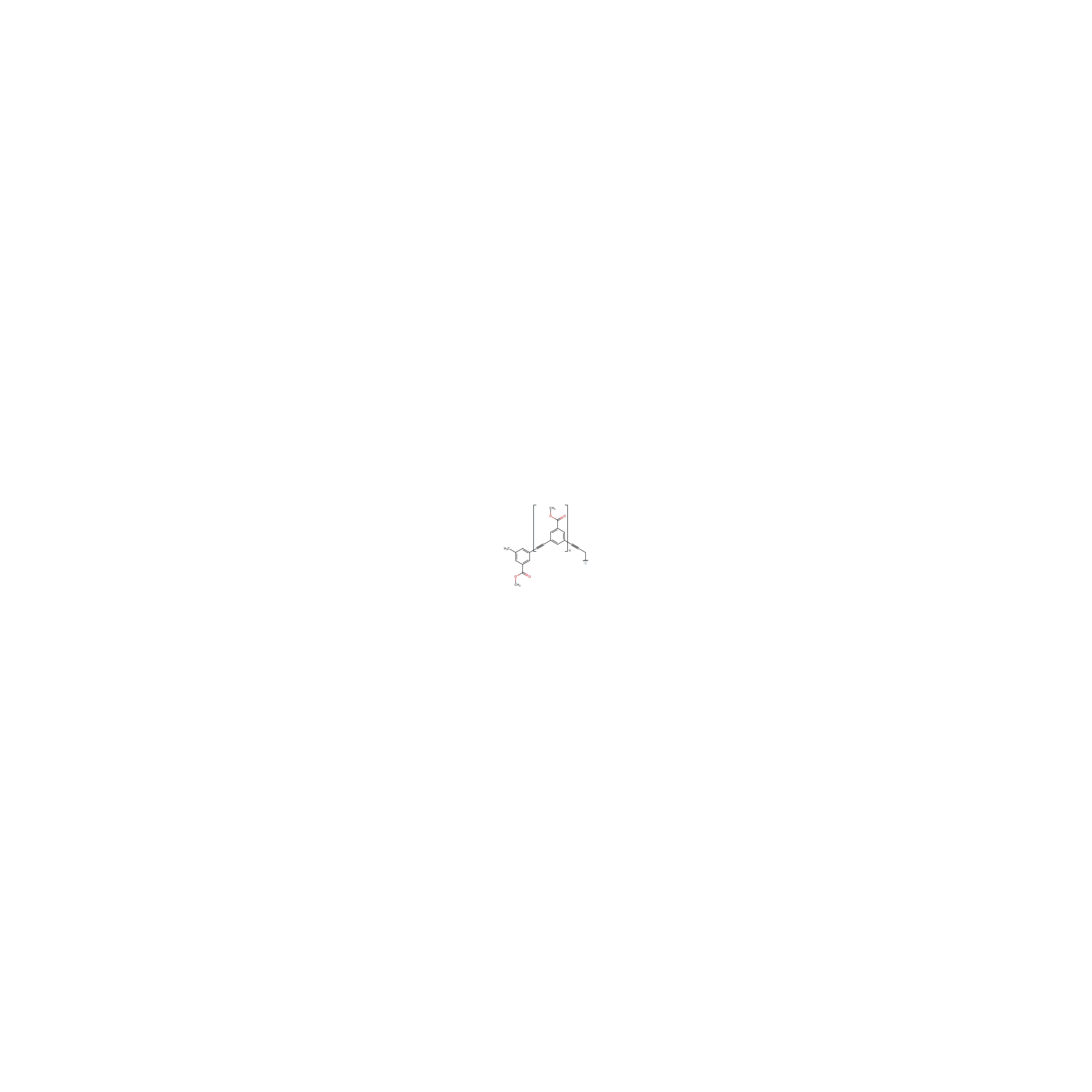}
\includegraphics[width=5in, trim=2cm 12cm 2cm 11cm, clip=true]{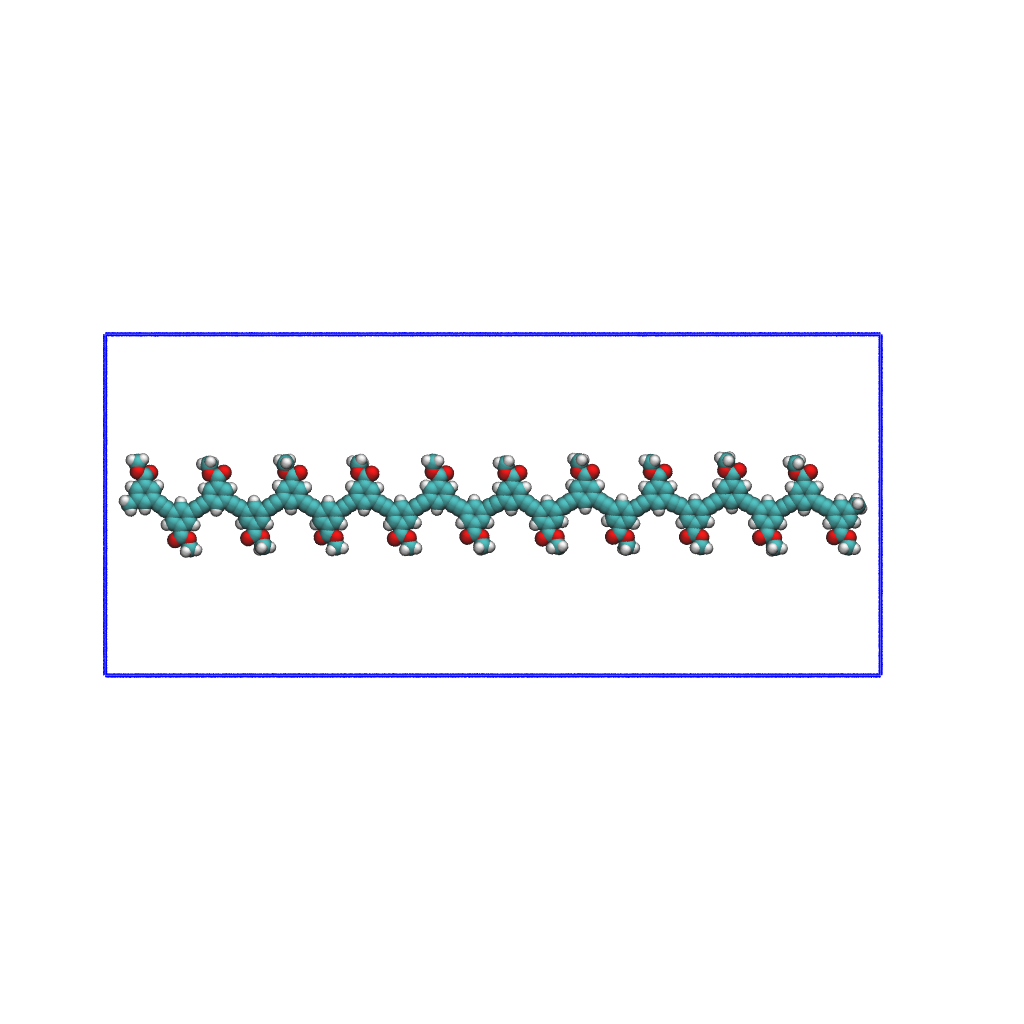}
\caption{(Top) Chemical structure of the \textit{meta}-substituted methylbenzoate phenylacetylene monomeric unit. (Bottom) Structureless elongated random coil view of the starting \textit{poly}-phenylacetylene (pPA) foldamer considered in this work. The structure was drawn with \textit{OpenBabel} online tool \cite{OBoyle2011}. The simulations were performed for three values of \textit{n} : \textit{n=12} dodecamer \textit{12mer}- ; \textit{n=16} hexadecamer \textit{16mer}- and \textit{n=20} eicosamer \textit{20mer}- polymers. This latter is shown as case-illustration example.}
\label{Fig:initial_linear_20mer}
\end{figure*}

The initial structures of the poly-phenylacetylene (pPA) oligomers were prepared with the online tool \textit{OpenBabel} \cite{OBoyle2011}, see Fig. \ref{Fig:initial_linear_20mer}. Their parameters and atom types were obtained from the Amber-compatible parameters of \textit{Gaff2} using Antechamber module of AmberTools \cite{Junmei2004, Wang2006}. Subsequently, those parameters were converted to Gromacs-like format of amber99sb-ildn \cite{Kresten2010} force field using the \textit{acypipy.py} script \cite{SousaDaSilva2012}. The potential energy function of the latter force field then takes the functional form described in Eq. \ref{pot_function}. The charges optimization were computed using AM1-BCC charge model \cite{Jakalian2000, Jakalian2002}. 

The simulations were performed in three different solvents with different polarities, \textit{n}-hexane (\ce{nC6H14}), cyclohexane (\ce{cC6H12}) and water (\ce{H2O}), listed in increasing order of polarity (see Supplementary Material and further below). In all cases the solvent molecules were added into a rectangular box containing the solute moiety aligned along the \textit{x}-axis. The different computed systems and their corresponding unit box dimensions are shown in Table \ref{simulated_system}.
While TIP3P water model \cite{Jorgensen1983} was used for simulations in aqueous milieu, a united atom representation for cyclohexane \ce{cC6H12} was prepared following our previous methodology \cite{Dongmo2020, Dongmo2023}. Meanwhile, the topology of \textit{n}-hexane \ce{nC6H14} was built as described above for the \textit{pPA}-oligomer. 

The final systems include atom numbers ranging from  15240 for the smallest system size, the \textit{12mer-pPA} in \textit{n}-hexane \ce{nC6H14}, to 83664 for the largest one, the \textit{20mer-pPA} in water \ce{H2O}, see Table \ref{simulated_system}. The solvent molecules were added to achieve the densities of about 970 g/L, 771 g/L, and 650 g/L for water \ce{H2O}, cyclohexane \ce{cC6H12} and \textit{n}-hexane \ce{nC6H14} respectively, roughly corresponding to the liquid phase density for each of the considered solvent at 300 K. 
\textcolor{black}{To test for possible finite size effects and yet keep the computational effort within reasonable limits, we have also performed additional simulations for the 12mers in water (box size $>2$ times the original box), and for 20mers in n-hexane (box size $>3$ times the original box. The results are reported in Supplementary Material and confirm that finite size effects do not hamper the present findings.}

Throughout the study, the atomistic molecular dynamics simulations (MD) were performed with Gromacs (version 2022.3, ) molecular package \cite{Plimpton1995}.  

\subsection{All-atom simulation details}
\label{subsec:simulations}
The solute’s potential energy of the solvated systems was minimized by relaxing the solvent around the solute atoms before running the unrestrained MD simulations. During the energy minimization stage, we employed the steepest descent minimization algorithm with a minimization step size of 0.01 nm and a maximum convergence force of 500.0 kJ mol$^{-1}$nm$^{-1}$. Thereafter, an equilibration round in the canonical \textit{NVT} ensemble was performed for 5 ns using the accurate leap-frog stochastic dynamics integrator with a simulation time step of 1 fs. 
While long-range electrostatics interactions were accounted with the Particle Mesh Ewald summation, short-range electrostatics and van der Waals interactions were truncated with a single-range cutoff at 12 \AA~with the pair list updated every 20 steps. The velocity for the Maxwell distribution temperature was set to 300 K. The temperature of the full system was equilibrated to this latter reference value using the velocity rescaling (modified Berendsen thermostat) \cite{Bussi2007} with a coupling constant of 0.1 ps. To mimic the density of the realistic bulk-like phase, all the simulations were replicated in the 3D space using periodic boundary conditions and all bonds involving hydrogen atoms were restrained using LINCS algorithms \cite{Hess1997}. 

The second equilibration run lasts 5 ns and was performed in the isobaric-isothermal \textit{NPT} ensemble using the same parameters as described above for \textit{NVT}. Moreover, the pressure was kept around the reference value of 1 bar using the Parrinello-Rahman pressure coupling \cite{Parrinello1981}  with a coupling constant of 1.0 ps.  In the final production stage, the restrains on heavy atoms were released and the systems evolved for 100 ns without imposing any constraints on solute’s bending and dihedral degrees of freedom, see Table \ref{simulated_system}. 

To improve the sampling statistics, five independent runs were submitted starting from the same initial structure, but using different random seed generators. 

In this description, the total potential energy of the system is given by
\begin{eqnarray}
\label{pot_function}
U&=&\sum_{bonds}\frac{1}{2}k_r\left(l_{i}-l_{i,0}\right)^2+\sum_{angles}\frac{1}{2}k_{\Theta}\left(\Theta_{i}-\Theta_{i,0}\right)^2 \nonumber \\
&+&\sum_{dihedrals}\frac{1}{2}k_\psi\left[1+\cos\left(n\psi+\Upsilon\right)\right] \nonumber \\
&+& \sum_{impropers}\frac{1}{2}k_{\omega}(\omega-\omega_0)^2 \nonumber \\
&+& \sum_{i<j}4\epsilon_{ij}\left[\left(\frac{\sigma_{ij}}{r_{ij}} \right)^{12}-\left(\frac{\sigma_{ij}}{r_{ij}}\right)^6 \right] \nonumber \\
&+&\sum_{i<j}\frac{q_iq_j}{r_{ij}} 
\end{eqnarray}

In Eq. \ref{pot_function}, the first four components are bonded potentials that describe bonds, angles, proper dihedrals and improper torsions of the covalent structure, respectively. The last two terms run over all pairwise atoms \textit{i} and \textit{j} separated from each other by a distance $r_{ij}= \lvert \mathbf{r}_j-\mathbf{r}_i\rvert$ and showcase the nonbonded interactions. The bond stretching and bond bending ($1^\textrm{st}$ and $2^\textrm{nd}$ terms in Eq.(\ref{pot_function})) model the energetic change accompanying the deformations of the bond lengths $l$ and bond angles $\Theta$ from their respective equilibrium values $l_0$ and $\Theta_0$. These two interactions are mainly computed using a harmonic-like potential with force constants $k_r$ and $k_\Theta$. The third and fourth terms in Eq.(\ref{pot_function}) account for bond rotations. Here $k_\psi$ is the height of the rotational barrier associated to the proper dihedral angle $\psi$ characterized by the torsional angle phase $\Upsilon$ for each Fourier component $n$ (periodicity). Improper torsions ensure planarity in aromatic rings and allow distinguishing molecules from their mirror images in which $k_\omega$ is the force constant for the improper dihedral $\omega$ going up and down its equilibrium position $\omega_0$. The last two components of Eq. \ref{pot_function} describe the van der Waals repulsive (at short distance, $r^{\textrm{-12}}$ term) and attractive (at long distance, $r^{\textrm{-6}}$ term) pairwise atomic forces between \textit{i} and \textit{j} shown as Lennard-Jones 12-6 potential, and the electrostatic interactions. The variables $q_i$ and $q_j$ are the partial charges on pairwise atoms \textit{i} and \textit{j} separated by the distance $r_{ij}$, $\sigma_{ij}$ is the distance at which the Lennard-Jones potential is zero and $\epsilon_{ij}$ is the well depth.

\begin{table*}
\centering
\caption{Summary of simulated systems. Five different runs were performed for each system starting from the same initial conformer but using different random seed.}
\begin{tabular}{|c|c|c|c|c|c|}
\hline
Solvents                     & \textit{pPA}-oligomer   & Simulation box($\textrm{nm}^3$)     & \#atoms & conc. (g/L) & time (ns) \\
\hline\hline
\multirow{3}{*}{\ce{H2O}}    & \textit{12mer} & 9.2$\times$5.5$\times$9.2   & 44988   & 965.71      & 100       \\   
                             & \textit{16mer} & 10.5$\times$5.5$\times$10.5 & 58545   & 969.71      & 100       \\   
                             & \textit{20mer} & 12.5$\times$5.5$\times$12.5 & 83644   & 972.67      & 100       \\   
\hline\hline
\multirow{3}{*}{\ce{cC6H12}} & \textit{12mer} & 9.2$\times$5.5$\times$9.2   & 18060   & 771.02      & 100       \\   
                             & \textit{16mer} & 10.5$\times$5.5$\times$10.5 & 23526   & 771.07      & 100       \\   
                             & \textit{20mer} & 12.5$\times$5.5$\times$12.5 & 33324   & 771.01      & 100       \\   
\hline\hline
\multirow{3}{*}{\ce{nC6H14}} & \textit{12mer} & 9.2$\times$5.5$\times$9.2   & 15240   & 650.29      & 100       \\  
                             & \textit{16mer} & 10.5$\times$5.5$\times$10.5 & 19854   & 620.00      & 100       \\   
                             & \textit{20mer} & 12.5$\times$5.5$\times$12.5 & 28110   & 650.21      & 100       \\   
\hline
\end{tabular}  
\label{simulated_system}
\end{table*}


\subsection{Potentials of Mean force (PMF)}
\label{subsec:potential}
In order to scrutinize the interaction modes when two cyclohexane \ce{cC6H12} or n-hexane \ce{nC6H14} entities approach each other, thereby characterizing the extent of their hydrophobicity, we computed the potential of mean force (PMF), \textit{W(r)} of each of these moieties. The set-up employed follows the methodology described by Sarma and Paul \cite{Sarma2011, Sarma2012} in which a system comprising 10 molecules of each type, parameterized in a united-atom like model representation, were randomly inserted into a cubic box of 15.24 $\textrm{nm}^3$ volume. Subsequently, 490 TIP3P water solvents were added to fill the simulation box, leading to an overall 500 molecules for each starting system, corresponding to a concentration of about 1.089 M ($\sim$ 0.013 g/L). After a preliminary steepest descent minimization, one round of \textit{NPT} equilibration with position restrains was performed for 10 ns using the Parrinello-Rahman pressure coupling ($\tau P$=0.5 ps). This run ensures a mechanical equilibration while the volume is fluctuating. The final box volume corresponding to the pressure of 1.01325 bar is then stabilized at the end of the simulation and used in the forthcoming runs. Thereafter, while still keeping the solute's atoms frozen, we performed a short \textit{NVT} equilibration for 10 ns using the velocity rescaling thermostat ($\tau T$=0.1 ps) thereby maintaining the temperature around 298.15 K. Finally, fully unrestrained MD runs in canonical \textit{NVT} were performed for 100 ns and the frames were saved every 25 ps. In all the simulations, a time-step of $10^{-15}$ s was used. 

The Lennard-Jones parameters $\sigma_{ij}$ and  $\epsilon_{ij}$ for two interacting sites \textit{i} and \textit{j} were obtained by employing the Lorentz-Berthelot combining rules $\sigma_{ij}=(\sigma_i+\sigma_j)/2$ and $\epsilon_{ij}=\sqrt{\epsilon_i\epsilon_j}$. Those values were taken equal to 0.3497 nm and 0.7266 KJ$\textrm{mol}^{-1}$ for $-$CH$_\textrm{2}$$-$ in cyclohexane \ce{cC6H12} \cite{Mauricio2015}, whereas the corresponding values for \textit{n}-butane as previously reported by Jorgensen et al. \cite{Jorgensen1984} were used for \ce{nC6H14} : 0.3905 nm and 0.7322 KJ$\textrm{mol}^{-1}$ for sp$^\textrm{3}$-methyl group, and 0.3905 nm and 0.4937 KJ$\textrm{mol}^{-1}$ for sp$^\textrm{3}$-methylene building units. Moreover, long-range electrostatics interactions were computed with the Particle Mesh Ewald scheme, while short-range electrostatics and van der Waals interactions were truncated with
a single-range cutoff at 12.2 Å with the pair list updated every 10 steps.  

The pair radial distribution functions \textit{g(r)} were computed for each of the molecular pairs involved, i.e.  \ce{cC6H12}-\ce{cC6H12} or \ce{nC6H14}-\ce{nC6H14}, \ce{cC6H12}-\ce{H2O} or \ce{nC6H14}-\ce{H2O}, and \ce{H2O}-\ce{H2O}. Subsequently, the PMF, $W(r)$ was computed using the relation 

\begin{eqnarray}
  \label{pmf}
 W(r) &=& -k_\textrm{B}T \ln g(r)
\end{eqnarray}
where $k_B T\approx $ 0.593 KJ$\textrm{mol}^{-1}$ at $T$=298.15 K.

\subsection{Solvation free energy}
\label{subsec:solvation}
The solvation free energy $\Delta G_{s}$ can be defined as the difference between the free energy of a solute in a specified solvent $G_{s}$ and in vacuum $G_{o}$
\begin{eqnarray}
  \label{sec2:eq1}
     \Delta G_{s} &=& G_{s} - G_{0}
\end{eqnarray}
If $\Delta G_{s}<0$ the process is spontaneous indicating that solvation is favored. This concept can clearly be extended to the free energy transfer from solvent $s_1$ to solvent $s_2$ 
\begin{eqnarray}
  \label{sec2:eq2}
 \Delta \Delta G_{s_1>s_2} &=& \Delta G_{s_{1}} - \Delta G_{s_{2}}
\end{eqnarray}

From the numerical viewpoint, free energy differences can be conveniently computed by using thermodynamic integration \cite{Leach2001}
\begin{eqnarray}
  \label{sec2:eq3}
  \Delta G_{s} &=& \int_{0}^{1} d\lambda \left \langle \frac{\partial V\left(\mathbf{r};\lambda\right)}{\partial \lambda} \right \rangle_{\lambda}
\end{eqnarray}
where $V(\mathbf{r},\lambda)$ is the potential energy of the system as a function of the coordinate vector $\mathbf{r}$, and $0\le \lambda \le 1$ is a switching-on parameter allowing a gradual change from state $\lambda=0$, where the solute is fully interacting, to state $\lambda=1$ where it does not interact at all.  The average $\langle \ldots \rangle_{\lambda}$ in Eq.(\ref{sec2:eq3}) is the usual thermal average with potential $V(\mathbf{r},\lambda)=(1-\lambda) V(\mathbf{r},0)+ \lambda V(\mathbf{r},1) $ at a fixed value of $\lambda$. The $\lambda$ interval $[0,1]$ is partitioned
into a grid of small intervals, molecular dynamics simulations are performed for each value of $\lambda$
belonging to each interval, and the results are then integrated over all values of $\lambda$ to obtain the final free energy difference. In the present study, 21 lambda points for each simulated system were used.

The solvation free energy was computed for each of the three polymer sizes in both solvents considered here at the temperature of 300 K. Following our previous protocol \cite{Dongmo2020, Dongmo2023}, we kept the polymers stretched, thereby maximizing the number of solute-solvent contacts, by applying harmonic restrains at the \textit{meta}-substituted sp$^\textrm{3}$-methyl carbon atom end-points. Moreover, cyclohexane \ce{cC6H12} and \textit{n}-hexane \ce{nC6H14} were both modelled in their united atoms conformations while for water \ce{H2O} molecules the TIP3P model was employed.  A simulation time step of 1 fs (for \ce{cC6H12}) or 2 fs (for \ce{nC6H14} and \ce{H2O}) was generally used, depending on the relative stability of the system. The accurate leap-frog stochastic dynamics integrator was applied in all the simulations, with the Berendsen coupling pressure scheme for simulations in \ce{cC6H12} and Parrinello-Rahman analog for those in \ce{nC6H14} and \ce{H2O}.        

Performing this calculation for different temperatures, allows to single out the separate contribution of the solvation enthalpy $\Delta H_{s}$ and the entropy $\Delta S_{s}$ as in Ref. \cite{Dongmo2020,Dongmo2023}.
%
\section{Results and discussion}
\label{sec:results}
\subsection{Bead-spring polymer in a Lennard-Jones liquid} 
\label{subsec:bead-spring_results}
We start our discussion by investigating the conformational changes of a single polymer chain, represented by a Kremer-Grest bead-spring model \cite{Kremer1990} in a solvent formed by spheres interacting via a Lennard-Jones soft potential. The quality of the solvent is regulated via the ratio $\epsilon_{ms}/\epsilon$ of the monomer-solvent attraction $\epsilon_{ms}$ and the monomer-monomer and solvent-solvent attraction $\epsilon$ assumed to be equal. Hence, $\epsilon_{ms}/\epsilon<1$ represents a \textit{poor} solvent, $\epsilon_{ms}/\epsilon>1$ a \textit{good} solvent, and $\epsilon_{ms}/\epsilon=1$ a \textit{neutral} ($\theta$) solvent. In this latter case, the conformational properties of the chain depends on the temperature only, whereas in good and poor solvents, they depend on both solvent quality and temperature.

\begin{figure*}[htbp]
\centering
\begin{subfigure}[b]{0.31\textwidth}
    \centering
    \includegraphics[width=5.45cm]{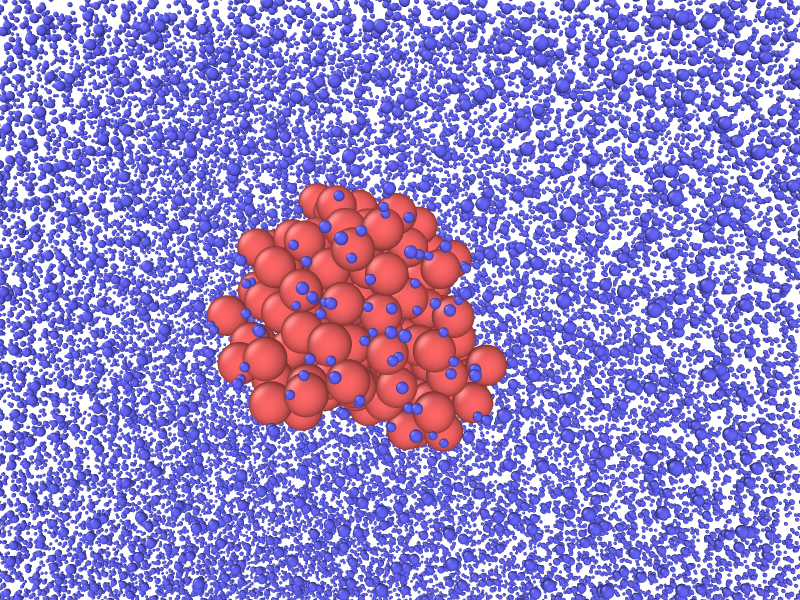}
     \caption{\label{fig:fig2a}}
    \end{subfigure}
    \quad
    \begin{subfigure}[b]{0.31\textwidth}
    \centering
    \includegraphics[width=5.45cm]{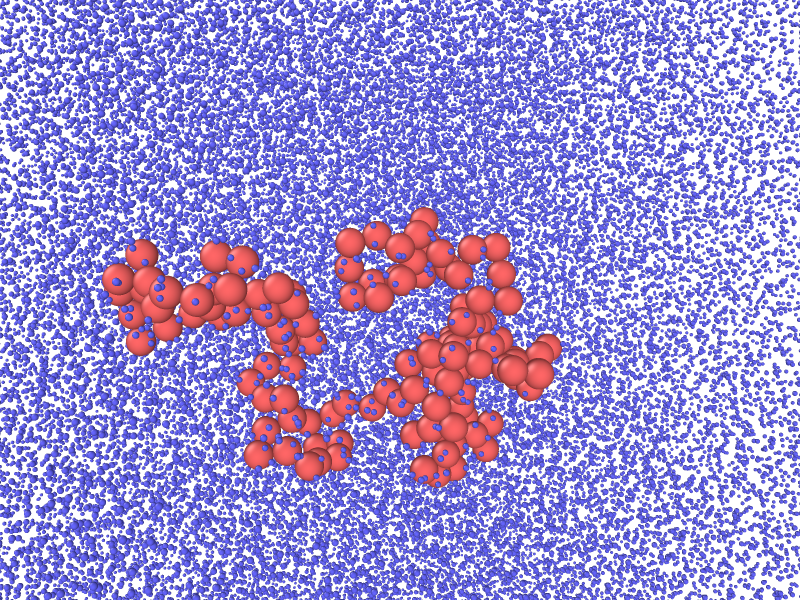}
     \caption{\label{fig:fig2b}}
    \end{subfigure}
    \quad
    \begin{subfigure}[b]{0.31\textwidth}
    \centering
    \includegraphics[width=5.45cm]{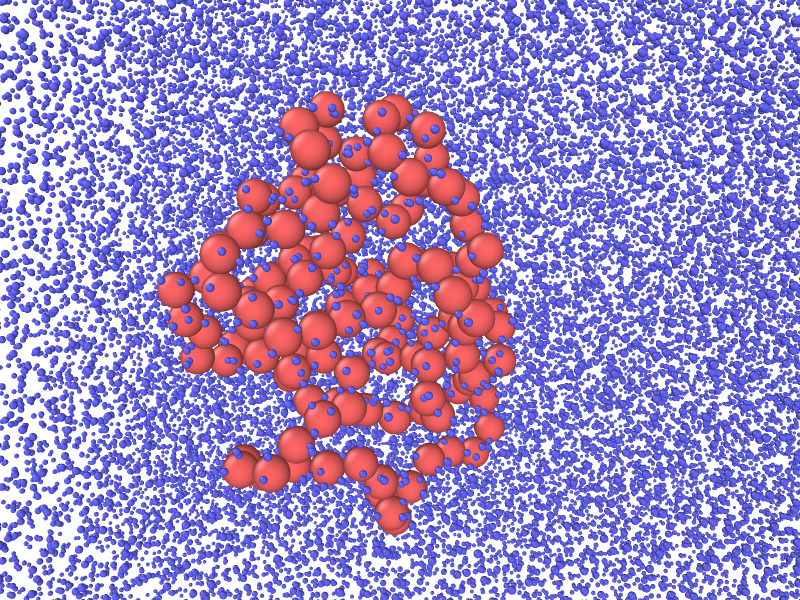}
     \caption{\label{fig:fig2c}}
    \end{subfigure}
    \begin{subfigure}[b]{0.31\textwidth}
    \centering
    \includegraphics[width=5.45cm]{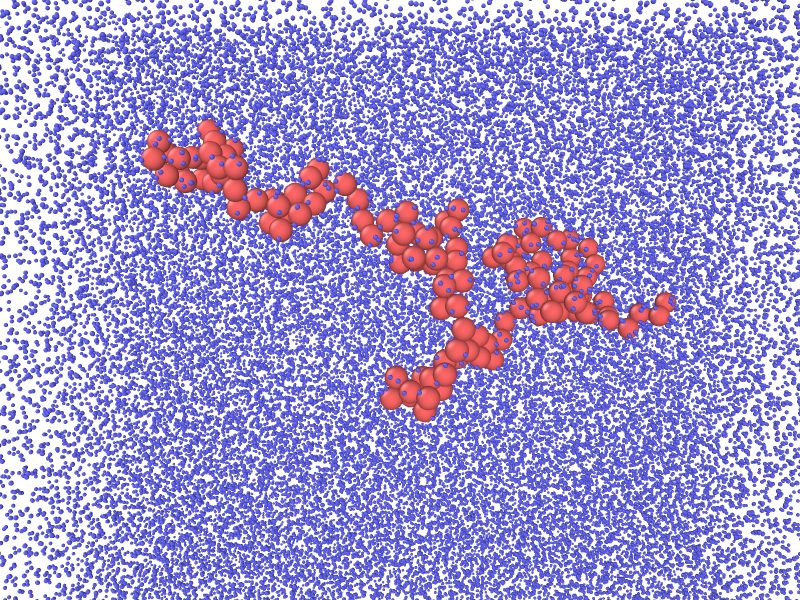}
     \caption{\label{fig:fig2d}}
    \end{subfigure}
    \quad
    \begin{subfigure}[b]{0.31\textwidth}
    \centering
    \includegraphics[width=5.45cm]{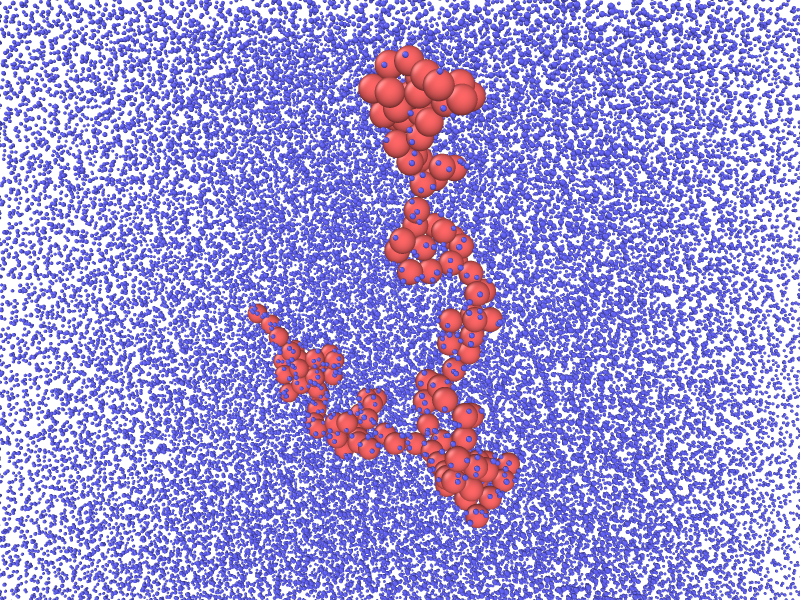}
     \caption{\label{fig:fig2e}}
    \end{subfigure}
    \quad
    \begin{subfigure}[b]{0.31\textwidth}
    \centering
    \includegraphics[width=5.45cm]{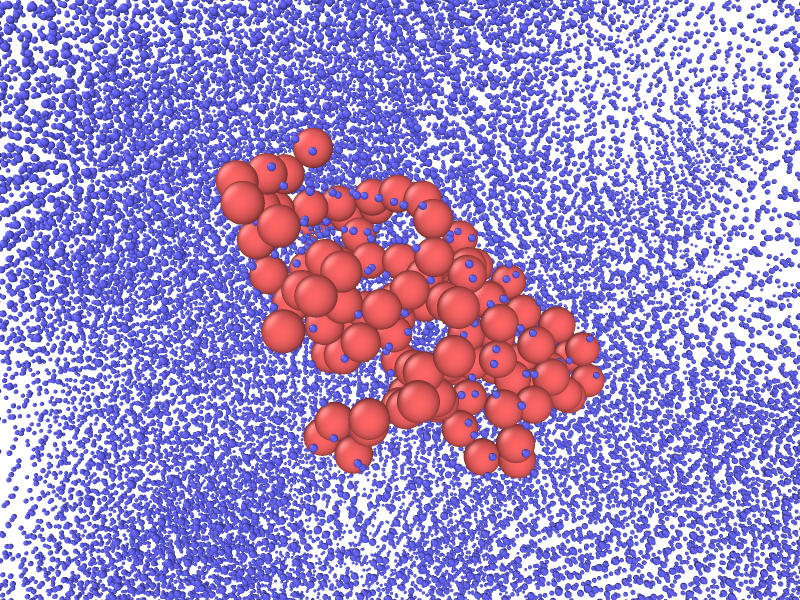}
     \caption{\label{fig:fig2f}}
    \end{subfigure}
  \caption{Snapshot of equilibrated configuration of a Kremer-Grest bead-spring polymer in a Lennard-Jones solvent at constant temperature $T^{*}=k_BT/\epsilon=1$ and different monomer-solvent interactions : (a) $\epsilon_{ms}/\epsilon=0.2$; (b) $\epsilon_{ms}/\epsilon=1.0$; (c) $\epsilon_{ms}/\epsilon=4.0$. Same as above but when the monomer-solvent interactions is $\epsilon_{ms}/\epsilon=1.0$ and for decreasing temperatures: (d) $T^{*}=2.0$; (e) $T^{*}=1.0$; (f) $T^{*}=0.1$. The condensation effect of the solvent in the latter case is particularly noteworthy. Cases (b) and (e) are statistically equivalent. The size of the solvent, nominally identical to the polymer bead, has been artificially reduced for clarity.}  
  \label{fig:lj_snapshots}
\end{figure*}

A detailed analysis of this paradigmatic system has already been carried out recently by \citeauthor{Huang2021}\cite{Huang2021}. Here we follow closely their approach, reproduce the part of their analysis that is of interest for the present study, and include an additional case that was not considered in Ref. \cite{Huang2021}.
Figure \ref{fig:lj_snapshots} depicts the collapsed conformation of the chain under (a) poor solvent condition $\epsilon_{ms}/\epsilon=0.2$, (b) neutral condition $\epsilon_{ms}/\epsilon=1.0$, and (c) good solvent condition $\epsilon_{ms}/\epsilon=4.0$. While in the first two cases, the results follow the expected behaviour of collapsing in poor solvent and of remaining swollen in neutral solvent, in the last case  $\epsilon_{ms}/\epsilon=4.0$ of strong monomer-solvent attraction, the chain is observed to undergo a collapse, albeit with a qualitatively different folded conformation compared to the poor solvent condition of case (a). The authors of Ref. \cite{Huang2021} ascribed this second collapse to a bridging mechanism occurring above a certain strength of the monomer-solvent attraction, where an effective monomer-monomer attraction sets in mediated by the solvent, mirroring a similar effect occurring in colloids \cite{Fantoni2015}. As the attraction increases from  poor solvent ($\epsilon_{ms}/\epsilon=0.2$), the radius of gyration $R_g$ (a standard order parameter for the onset  of the coil-globule transition) is observed to increase at the onset of the neutral ambient ($\epsilon_{ms}/\epsilon=1$), to remain swollen with large $R_g$ until  ($\epsilon_{ms}/\epsilon \approx 2.5$) and to collapse again above this value. In Ref. \cite{Huang2021} it was explicitly checked that both phases are collapsed phases with the correct $R_g \approx N^{1/3}$ behaviour. In this second folded conformation, however, some solvent molecules remains inside the polymer globule and induce an effective monomer-monomer attraction. This, along with the gain in solvent entropy achieved upon folding, stabilizes this folded conformation against the coil counterpart.
All simulations started with a preliminary equilibration of  $4.0 \times 10^6 \tau$ with the equilibrated solvent at a density of about $0.64 m/\sigma^3$ \cite{Huang2021}. To avoid possible surface effects, the number of solvent beads was adjusted so that the size of the cubic box was sufficiently larger than the radius of gyration.
\begin{figure*}[htbp]
\centering
\begin{tikzpicture}
\node[anchor=south west,inner sep=0] (image) at (0,0){\scalebox {0.85}{\includegraphics[width=0.8\textwidth, trim=0cm 0cm 4cm 2cm, clip=true, angle=0, page=1]{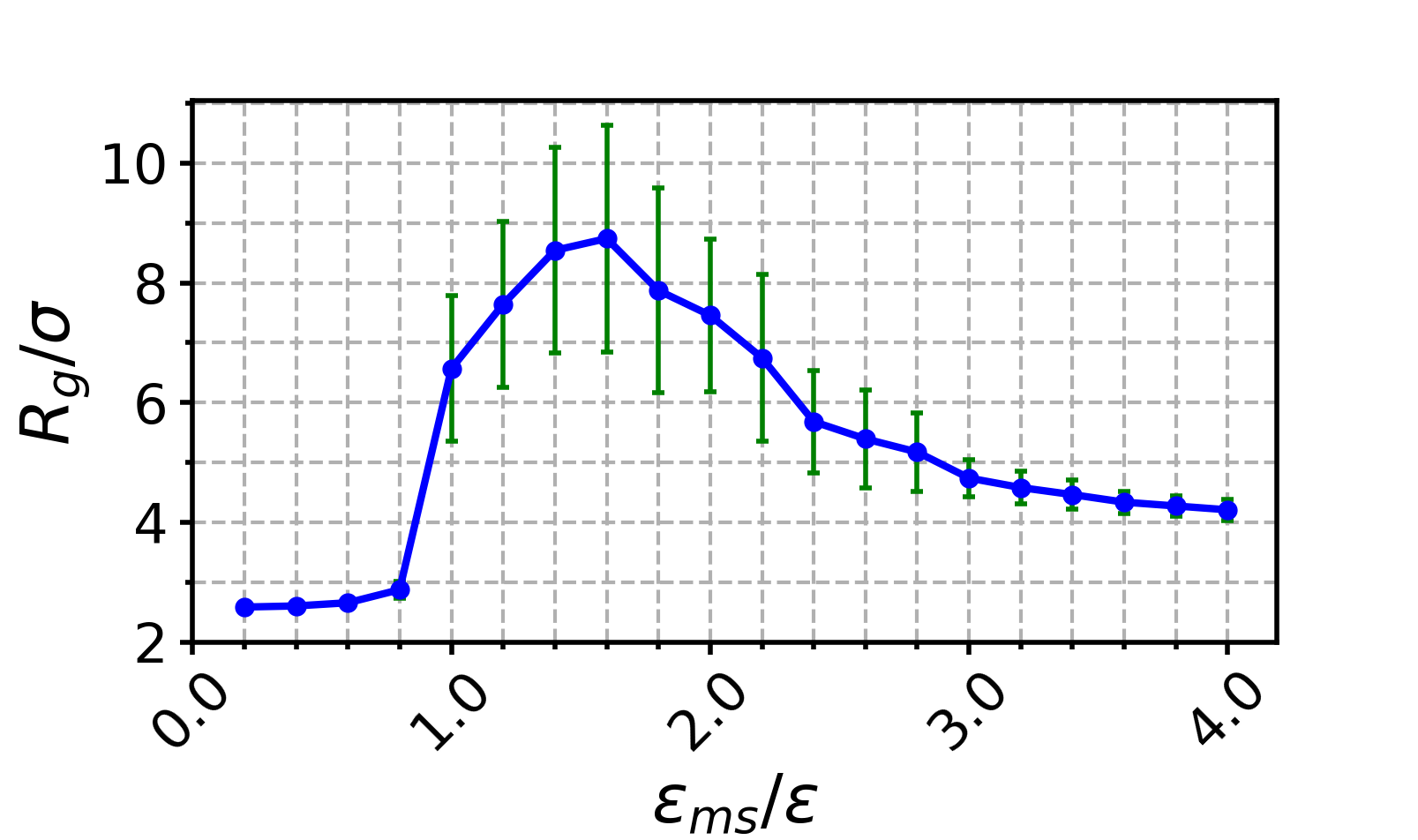}}};
\end{tikzpicture}
\begin{tikzpicture}
\node[anchor=south west,inner sep=0] (image) at (0,0){\scalebox {0.85}{\includegraphics[width=0.8\textwidth, trim=0cm 0cm 4.5cm 2cm, clip=true, angle=0, page=1]{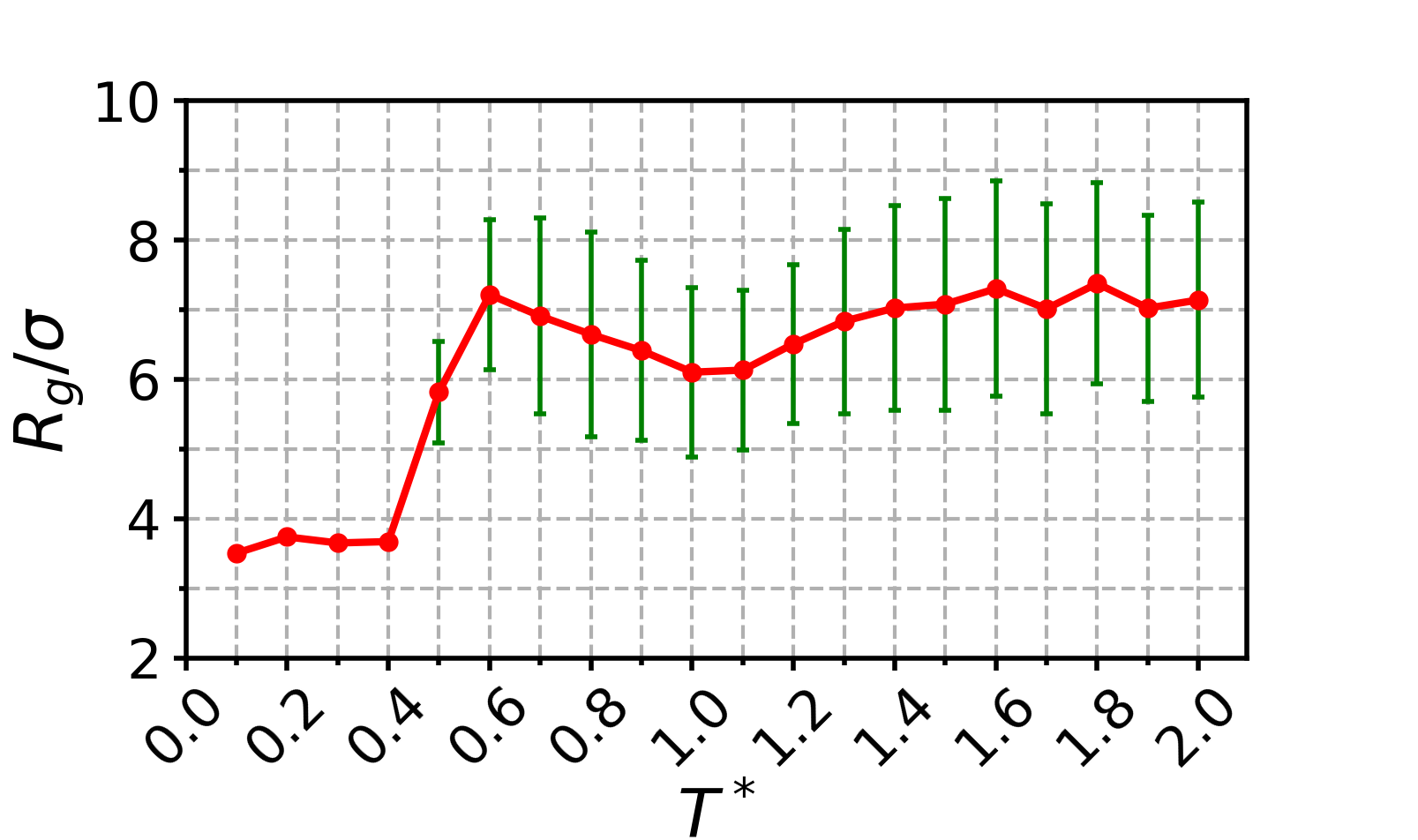}}};
\end{tikzpicture}
\caption{(Top) Radius of gyration $R_g$ (in units of $\sigma$) as a function of $\epsilon_{ms}/\epsilon$ in the case of $N=128$ monomers. In this case $T^{*}=1$ corresponding to room temperature; (Bottom) Radius of gyration $R_g$ (in units of $\sigma$) as a function of the reduced temperature $T^{*}$ again for $N=128$. In this case $\epsilon_{ms}/\epsilon=1$ corresponding to " neutral" ($\theta$) solvent. Error bars are displayed as vertical bars centered in each point.}
\label{fig:md_explicit}
\end{figure*}

Interestingly, the same re-entrant behaviour is \textit{not} observed in the case where a polymer collapses in a neutral solvent ($\epsilon_{ms}/\epsilon=1$) upon cooling the temperature. This case was not analyzed in Ref. \cite{Huang2021} and its reported here in Figures \ref{fig:fig2d}-\ref{fig:fig2f}, where we see that a chain originally swollen at $T^{*}=2.0$ (Figure \ref{fig:fig2d}) becomes slightly more compact under $\theta$ condition $T^{*}=1.0$, and eventually collapses into a globule upon further cooling at $T^{*}=0.1$. A quantitative measure of these conformational changes can be obtained from the radius of gyration $R_g$. 
Figure \ref{fig:md_explicit} (top panel) depicts the reduced radius of gyration $R_g/\sigma$ as a function of $\epsilon_{ms}/\epsilon$, a measure of the solvent quality, from the simulations with $N=128$ monomers. Here the reduced temperature is $T^{*}=1$ corresponding to room temperature, and this is the same case whose snapshots are reported in Figure \ref{fig:lj_snapshots}. At low $\epsilon_{ms}/\epsilon$ (poor solvent regime) the polymer is collapsed and $R_g/\sigma$ is very small. Upon approaching $\epsilon_{ms}/\epsilon=1$ (neutral solvent regime),  $R_g/\sigma$ starts to increase until reaching a maximum of $R_g/\sigma \approx 9$ at $\epsilon_{ms}/\epsilon=1.4$ (good solvent regime). A further increase of $\epsilon_{ms}/\epsilon$ leads to a marked decrease of $R_g/\sigma$ that is to be associated to a re-entrant collapse mediated by the solvent, as anticipated. Although, the details of the calculation are slightly different, these results exactly reproduce the findings of Ref. \cite{Huang2021}, supporting the robustness of this water-bridging interpretation. Figure \ref{fig:md_explicit} (bottom panel) reports the same calculation in which the solvent quality is kept fixed at neutral condition  $\epsilon_{ms}/\epsilon=1$, and temperature is reduced. Unlike previous case, clearly we observe a single coil-globule transition as a function of the (reduced) temperature $T^{*}$. It is important to remark that the solvent itself tends to collapse upon cooling (see Figure \ref{fig:fig2f}), and this explains the difference between this results and the textbook results from simulations in implicit solvents that show a much more marked transition. While it is rewarding to observe a consistent picture stemming from different calculations, this example highlights how the equivalence \textit{decreasing the temperature=decreasing the solvent quality} must be taken with great care, even in this very simple and paradigmatic example. Indeed, the actual phase diagram for this problem has been further shown to be even richer \cite{Garg2023}.
\subsection{Potential of mean force of cyclohexane \ce{cC6H12} and n-hexane \ce{nC6H14} in water}
\label{subsec:potentialMF}
\begin{figure*}[htbp]
\centering
\captionsetup{justification=raggedright,width=0.85\linewidth}
\begin{overpic}[scale=0.85, trim=0cm 0cm 0cm 0cm, clip=true, angle=0]{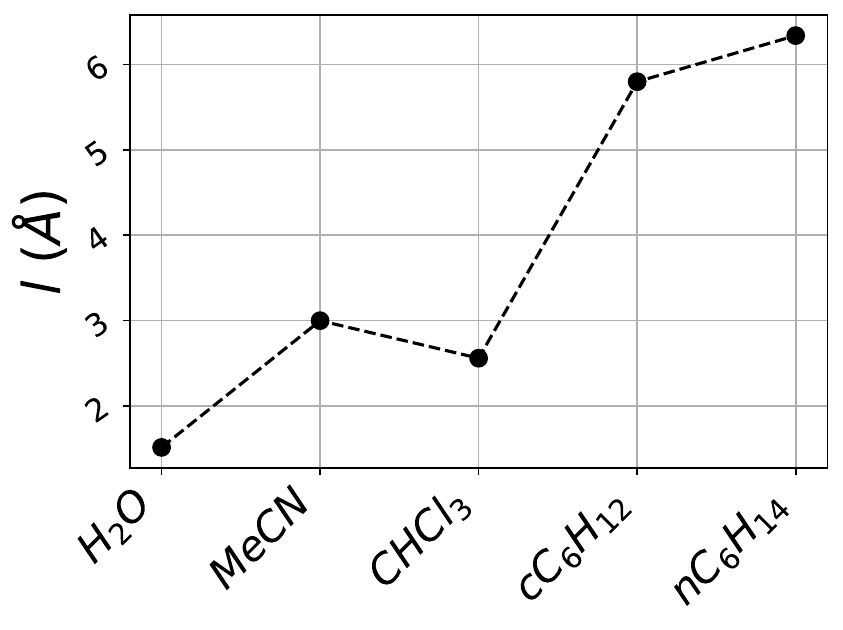}
       \put(13,23){\includegraphics[scale=0.1, trim=7.6cm 13.5cm 7.6cm 11.1cm, clip=true, angle=0]{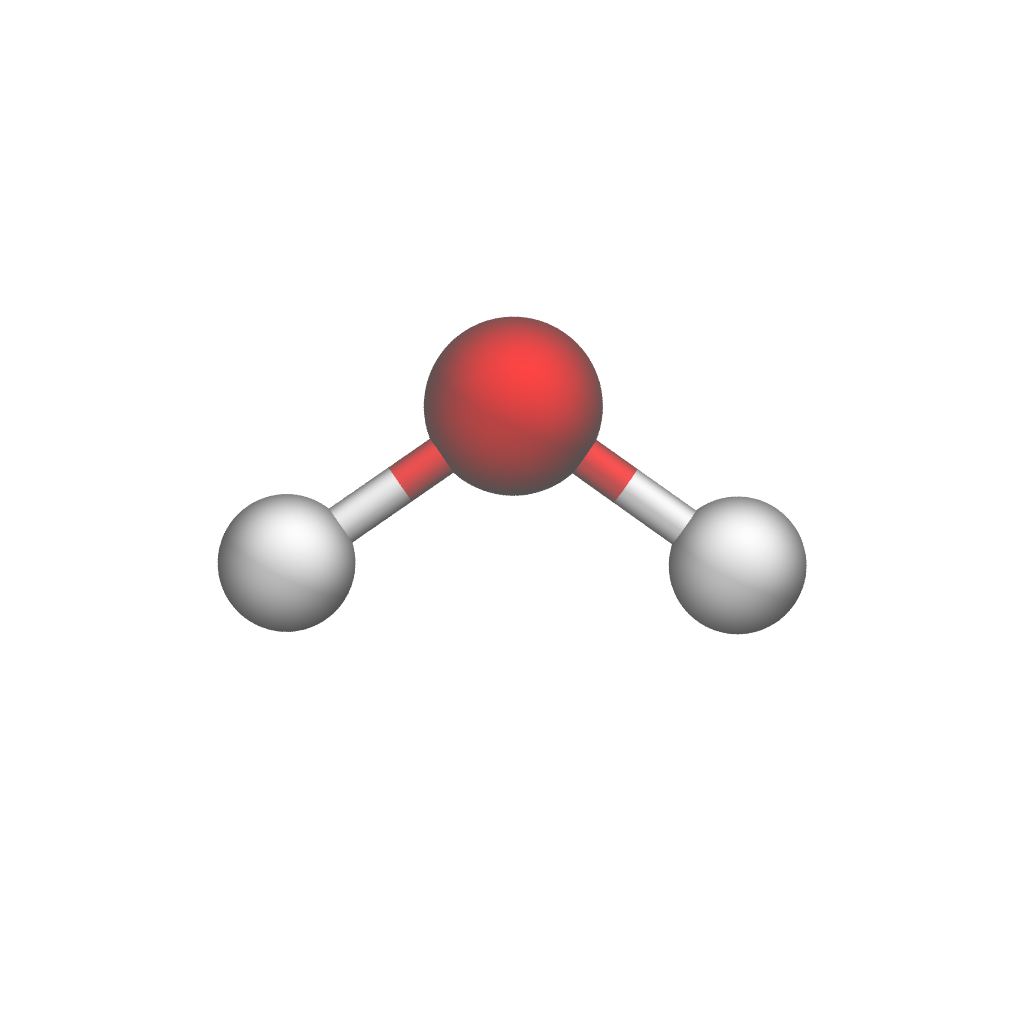}}  
       \put(54,65){\includegraphics[scale=0.075, trim=5.4cm 5.4cm 5.4cm 5.4cm, clip=true, angle=110]{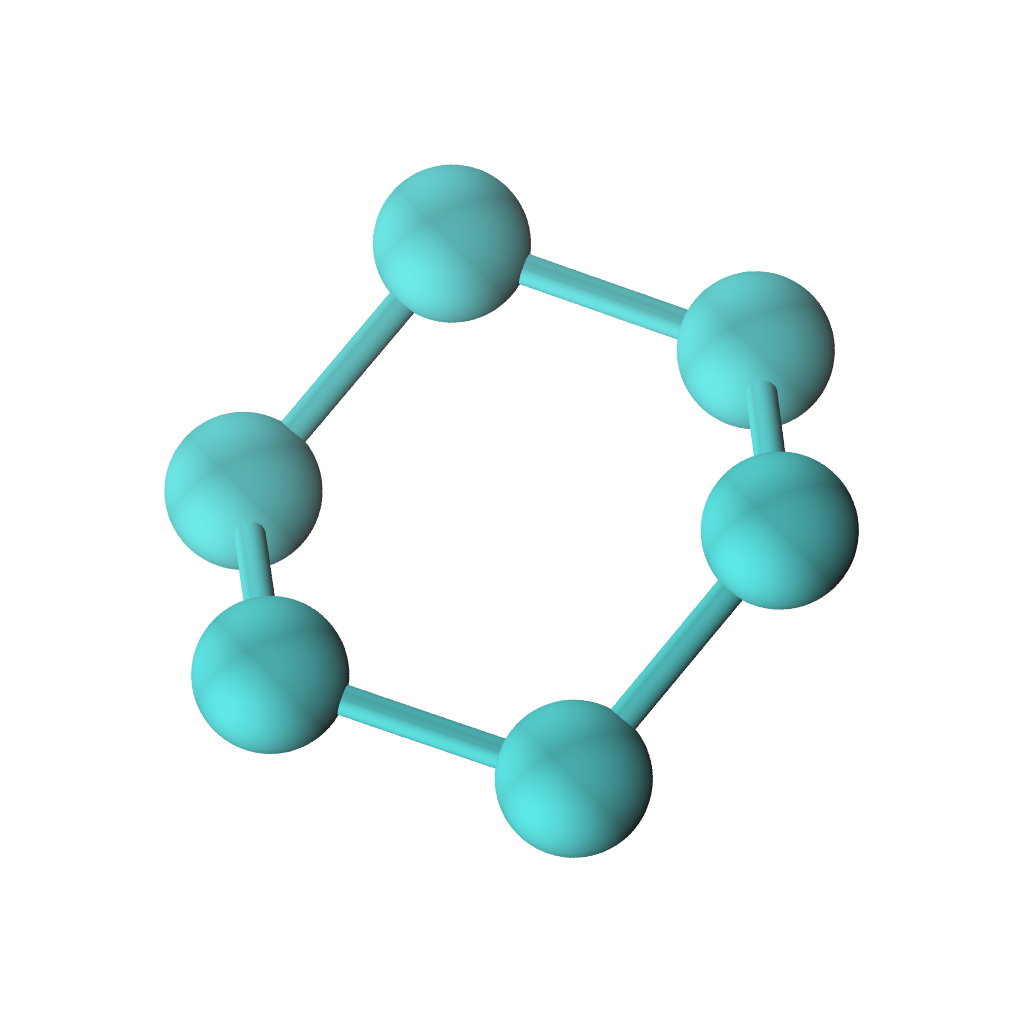}}
       \put(79,72){\includegraphics[scale=0.17, trim=6.5cm 15.1cm 6.5cm 15.1cm, clip=true, angle=0]{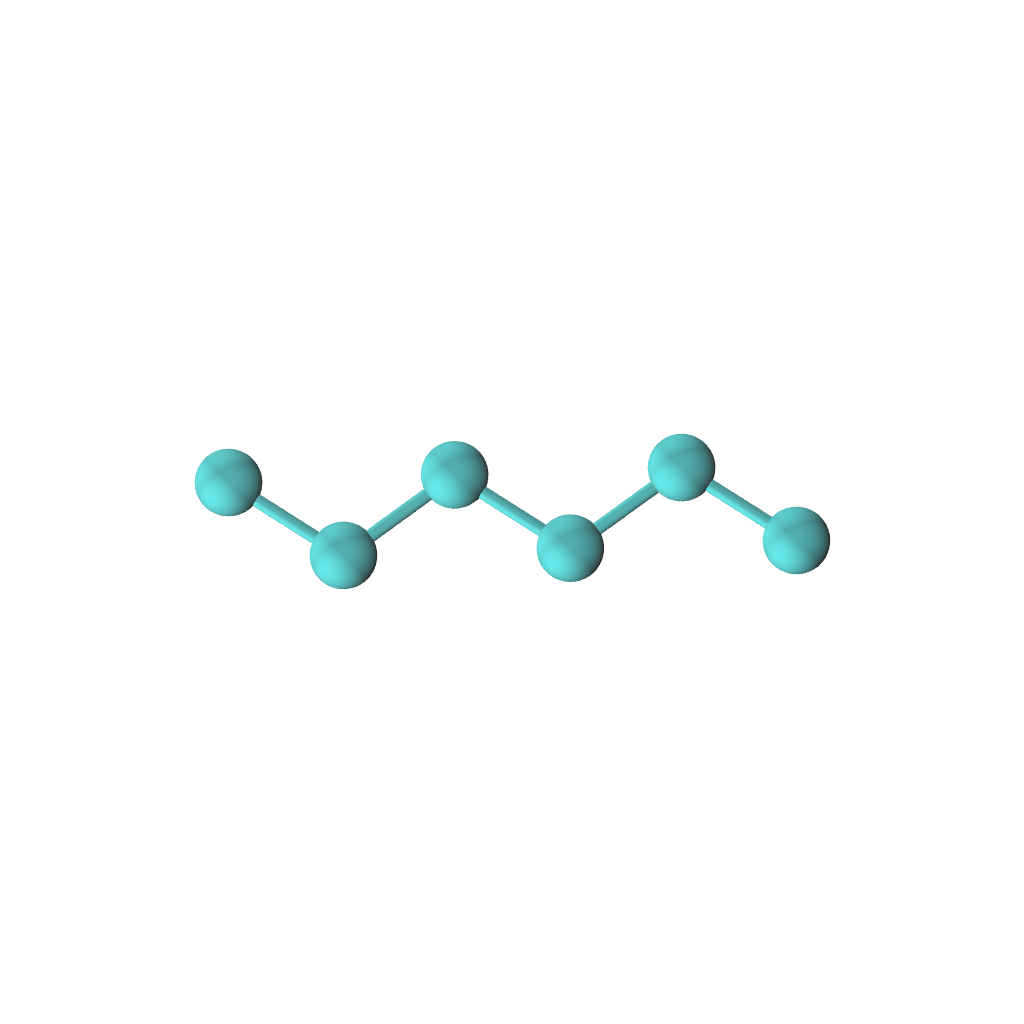}} 
       \put(19,37.5){\includegraphics[scale=0.13, trim=5.5cm 9cm 4.3cm 9.8cm, clip=true, angle=0]{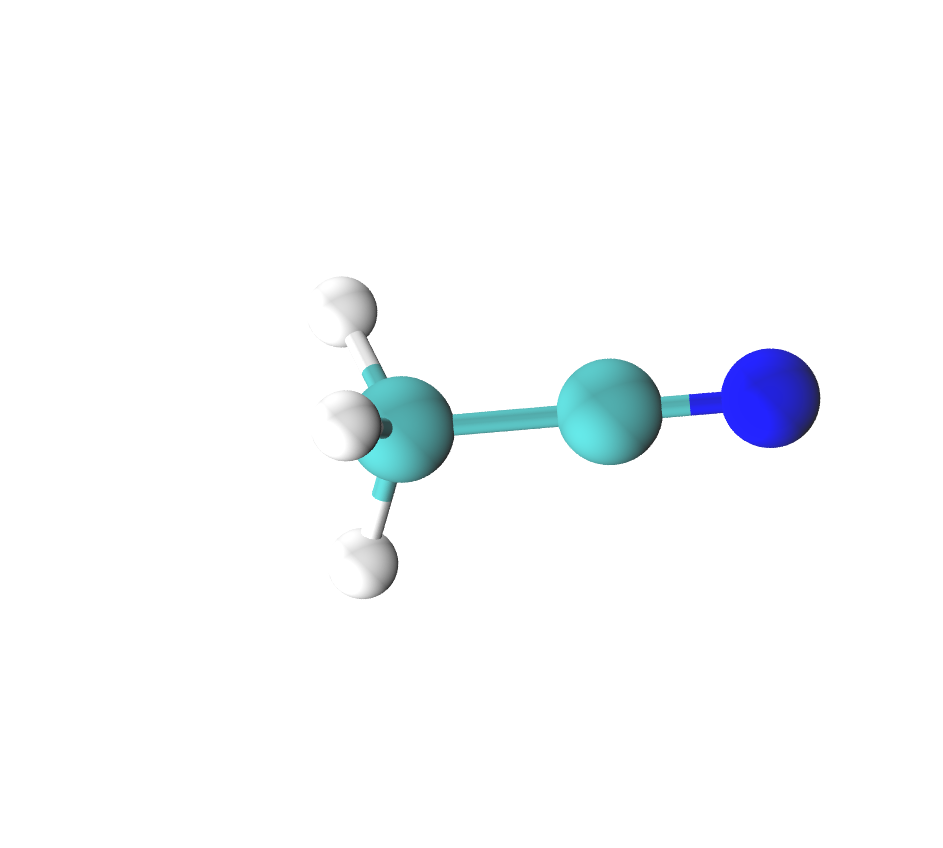}} 
       \put(49,34){\includegraphics[scale=0.12, trim=4.8cm 6.9cm 3.6cm 8.2cm, clip=true, angle=0]{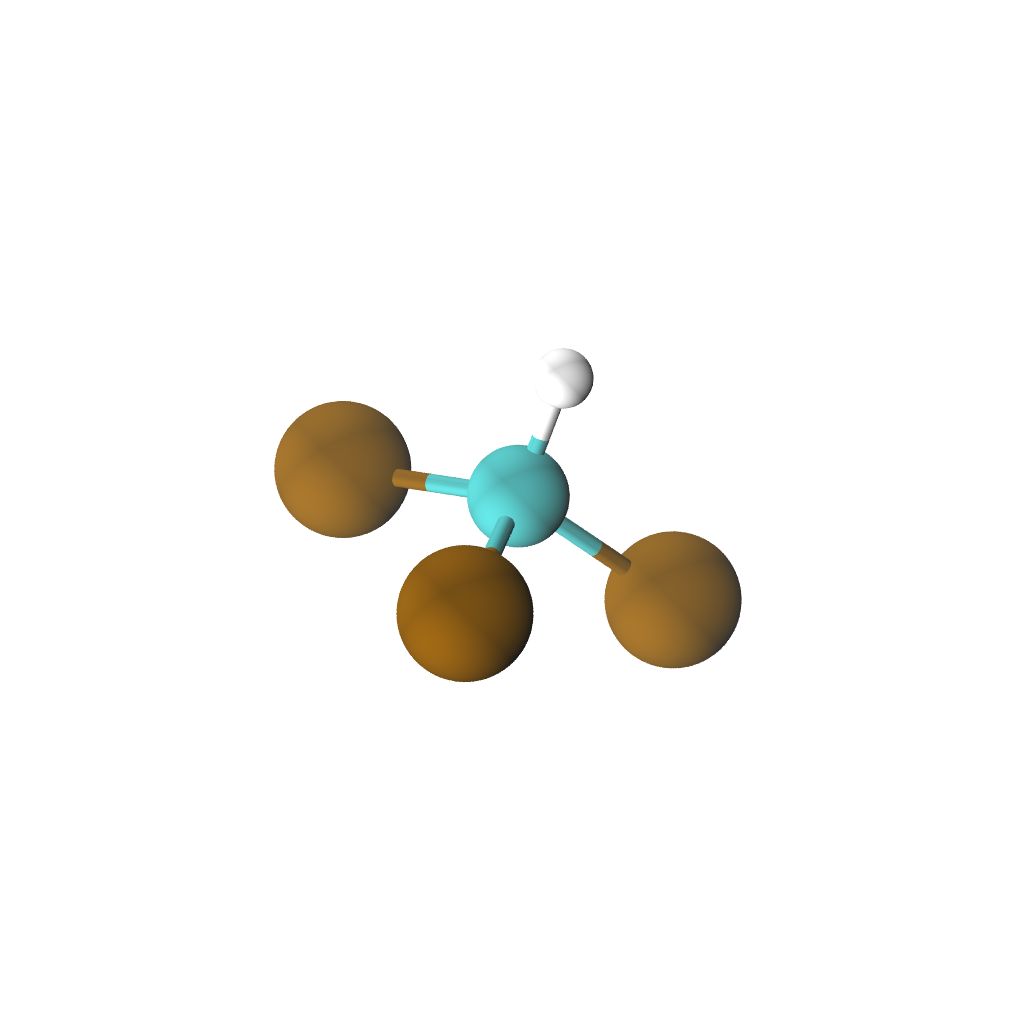}} 
    \end{overpic}
     \caption{Characteristic sizes $l$ (vertical axis) of the considered solvents as a function of their polarities (horizontal axis).  From left to right the overall size of water \ce{H2O} (\textbf{1.515\AA}), chloroform \ce{CHCl3} (\textbf{2.56\AA}), acetonitrile \ce{MeCN} (\textbf{3.00\AA}), cyclohexane \ce{cC6H12} (\textbf{5.80\AA}) and \textit{n}-hexane \ce{nC6H14} (\textbf{6.34\AA}), respectively. The length of water is easily computed from the O--H bond stretching ($\sim$0.943\AA) and H--O--H bending angle ($\sim$106$^\circ$). The length of cyclohexane is obtained from the work of Fomin and coworkers \cite{Fomin2015}, and that of \textit{n}-hexane estimated from Boese et \textit{al}. \cite{Boese1999}. Other data from Ref. 
     \textit{Herzberg, G., Electronic spectra and electronic structure of polyatomic molecules,Van Nostrand,New York, 1966}}
\label{fig:characteristic_lenght}
\end{figure*}

The analysis of previous Section  clearly shows that the conventional paradigm of good/poor solvent falls short in describing even a simple bead-spring model in a Lennard-Jones solvent. Consider now a real polymer in a real solvent, both of which will then be described atomistically. A polymer is customarily described in terms of structural units (the monomers) that are chemical moieties that can be either polar or hydrophobic. Likewise, the solvent can be either polar or hydrophobic or something in between. Strictly speaking, a polar molecule has a permanent dipole (hence the name), whereas a hydrophobic molecule has not. However, for a more complex chemical moiety -- in particular for a solvent, this distinction become fuzzy and one frequently used quantification of the polarity of the solvent is via the relative dielectric constant. Accordingly, a solvent is polar if the relative dielectric constant is high, hydrophobic if the relative dielectric constant is low. 
Three different solvents are considered in the present study.  Water \ce{H2O} is highly polar with a relative dielectric constant $80.1$.
 While liquid water anomalies still defy a complete description \cite{Gallo2016}, the geometry of the molecule is very simple with a typical length scale of approximately 1.5\AA and a H--O--H angle of $\approx 106^\circ$ (Figure \ref{fig:characteristic_lenght}).
 At the opposite side of the scale, there are the organic solvents with low relative dielectric constant.  Cyclohexane \ce{cC6H12} has relative dielectric constant $2.02$, a typical size of 5\AA, and is formed by a nearly flat single a 6-vertex aromatic ring  (Figure \ref{fig:characteristic_lenght}). Similarly, n-hexane \ce{nC6H14} is a straight-chain alkane with 6 carbon atoms with a characteristic size of 6 \AA   (Figure \ref{fig:characteristic_lenght}) and it has a relative dielectric constant $1.88$.
 For comparison, Fig. (\ref{fig:characteristic_lenght}) also includes  acetonitrile \ce{MeCN} (dielectric constant $37.5$) and chloroform \ce{CHCl3} (dielectric constant $4.81$) as polar and hydrophobic solvents already used in this framework \cite{Prince1999}. Hence, Figure (\ref{fig:characteristic_lenght}) is ordered according to the dielectric constants on the horizontal axis and according to the characteristic size $l$ on the vertical axis. 

\begin{figure}[htbp]
\centering
\begin{tikzpicture}
\node[anchor=south west,inner sep=0] (image) at (0,0){\scalebox {0.31}{\includegraphics[width=1.3\textwidth, trim=10cm 11.5cm 10.5cm 11cm, clip=true, angle=0, page=1]{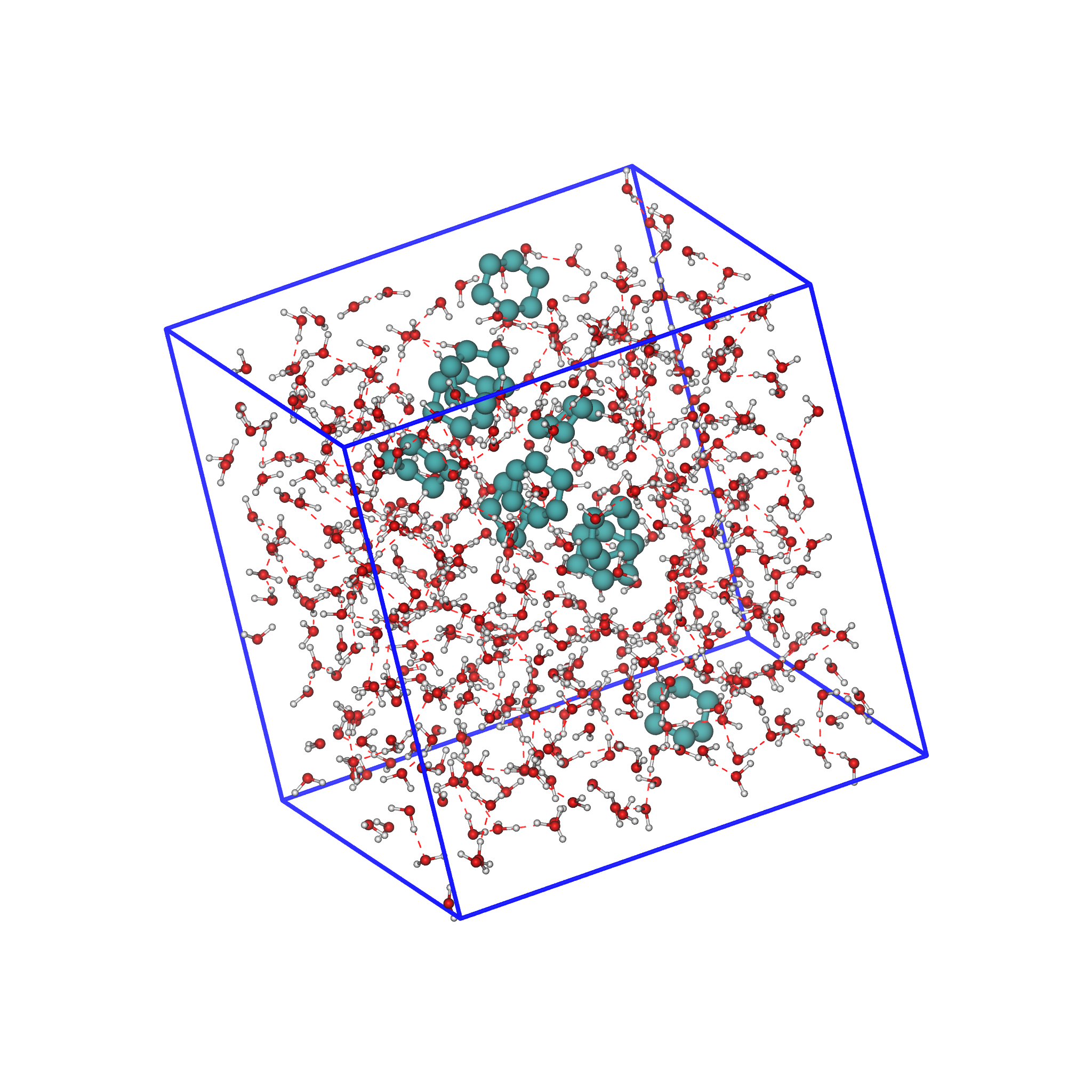}}};
\end{tikzpicture}
\begin{tikzpicture}
\node[anchor=south west,inner sep=0] (image) at (0,0){\scalebox {0.31}{\includegraphics[width=1.3\textwidth, trim=10cm 16cm 10.5cm 15cm, clip=true, angle=0, page=1]{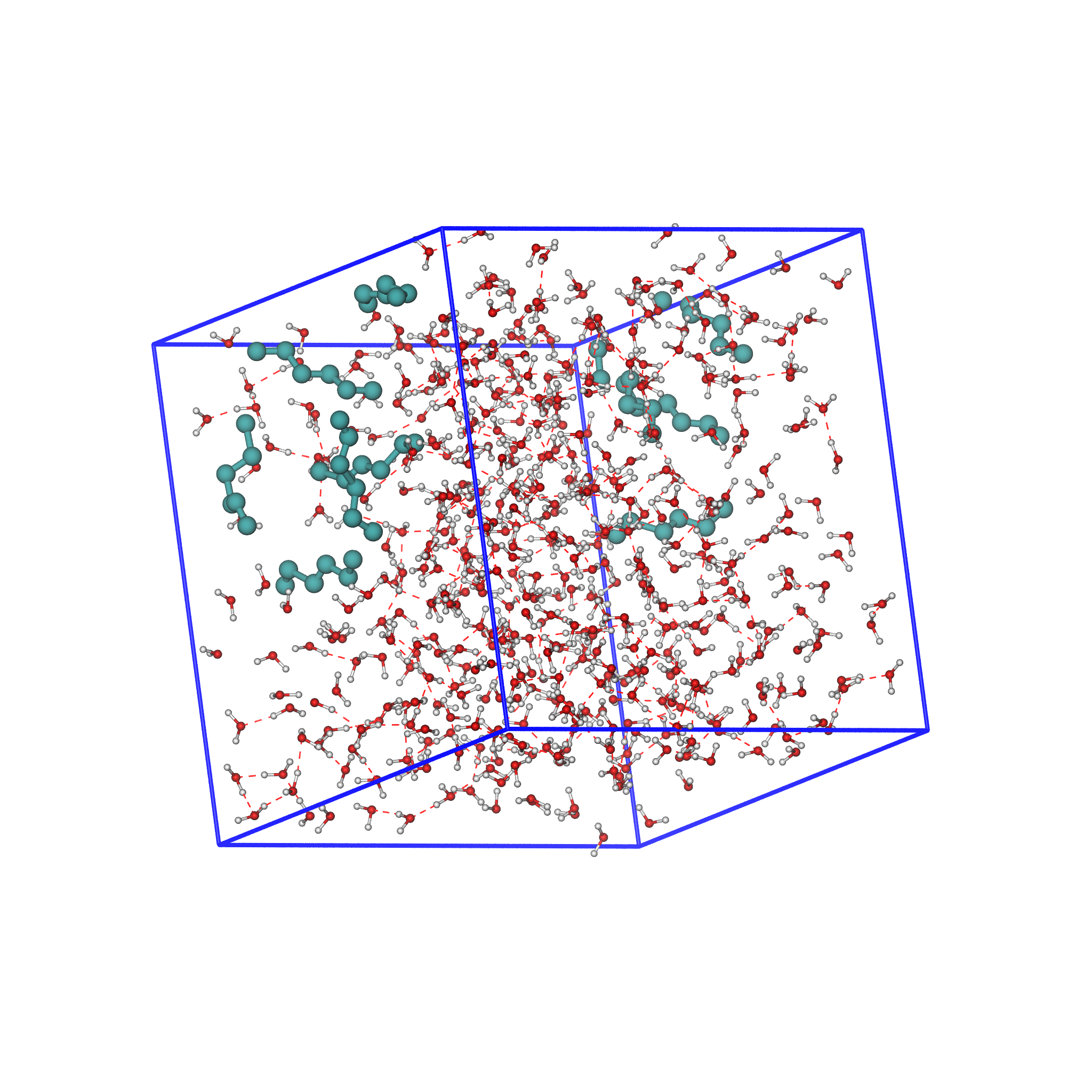}}};
\end{tikzpicture}
\caption{Snapshots of equilibrated cyclohexane \ce{cC6H12} in water \ce{H2O} (Left) and \textit{n}-hexane \ce{nC6H14} in water \ce{H2O} (Right). Hydrogen bonds forming network between solvent entities are explicitly displayed.}
\label{fig:hydrophobicity_snap}
\end{figure}
\begin{figure}[htbp]
\centering
\begin{tikzpicture}
\node[anchor=south west,inner sep=0] (image) at (0,0){\scalebox {0.85}{\includegraphics[width=0.6\textwidth, trim=0cm 0cm 0cm 0cm, clip=true, angle=0, page=1]{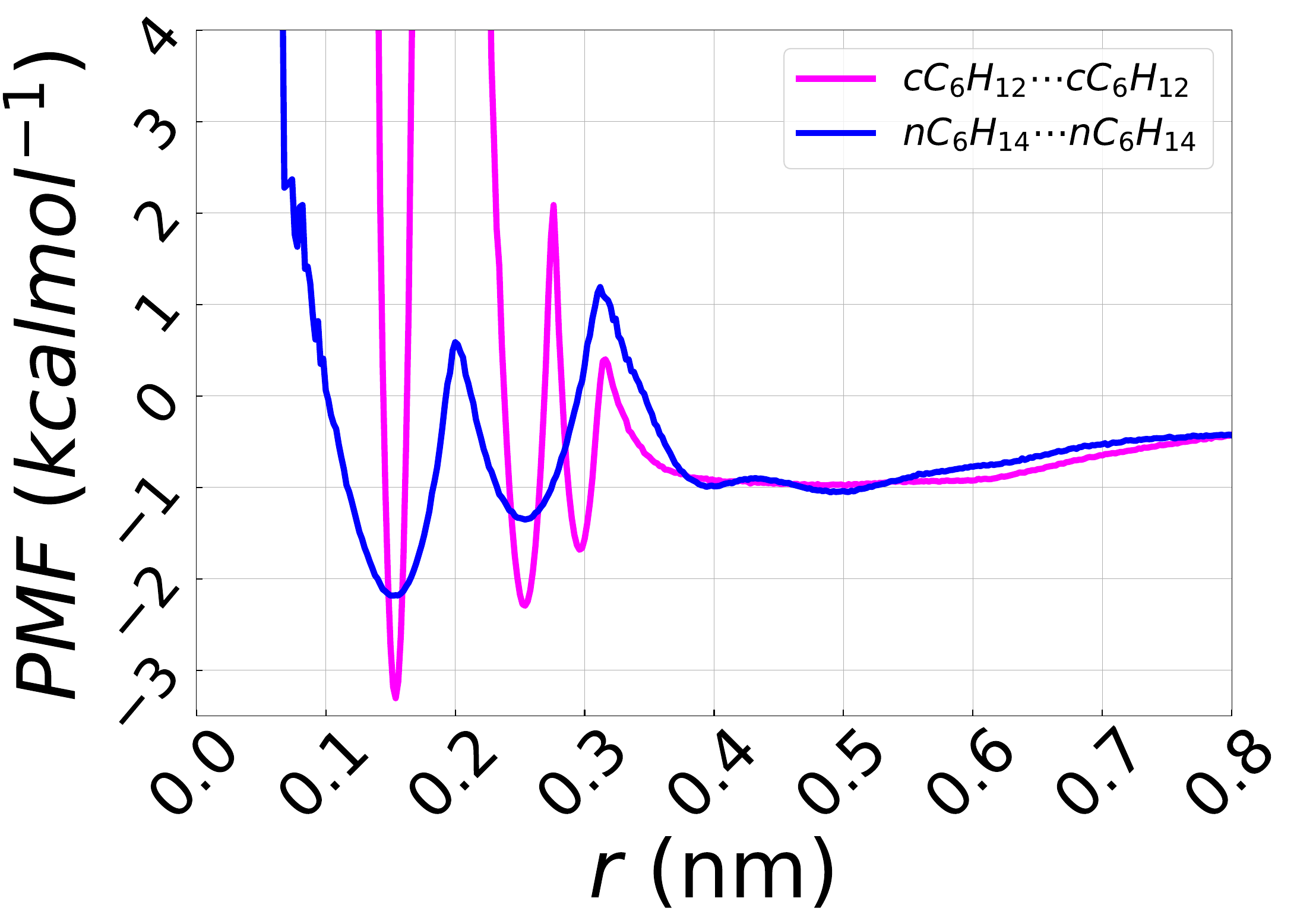}}};
\end{tikzpicture}
\caption{Potentials of mean force of cyclohexane \ce{cC6H12}  (magenta solid line) and of n-hexane \ce{nC6H14} (blue line) molecules in water \ce{H2O} solvent. The corresponding radial distribution functions are shown in Supplementary Material. The considered moieties are the center of the aromatic ring for cyclohexane and the central carbon for n-hexane (see text). Only non-bonded interactions were considered.}
\label{fig:pmf_hydrophobicity}
\end{figure}
Following Ref. \cite{Smith1993}, we performed molecular dynamic simulations of both cyclohexane \ce{cC6H12} and n-hexane \ce{nC6H14} in water to illustrate the hydrophobic effect, that is the tendency of hydrophobic solutes to aggregate to avoid unfavourable contacts with water \cite{BenNaim2012}. Figure \ref{fig:hydrophobicity_snap} reports the final configuration of 10  cyclohexane \ce{cC6H12} molecules (left panel) and 10 n-hexane \ce{nC6H14} moecules (right panel) in a aqueous solution formed by 490 water \ce{H2O} molecules, thus providing a visual confirmation of the tendency of both cyclohexane \ce{cC6H12} and n-hexane \ce{nC6H14} to aggregate in water \ce{H2O}. They do it differently, however. A quantitative way to assess the relative hydrophobicity of the two organic solvents \ce{cC6H12} and \ce{nC6H14} relies on the calculation the potential of mean force $W(r)$ between the two points distant $r$ apart in a water solution and belonging to two different molecules. In both cases we used the center-of-mass (COM) distance between pairs of molecular moieties to measure the distance $r$. In the case of cyclohexane \ce{cC6H12} COM is nearly coincident with the center of the aromatic ring, whereas in the case of n-hexane \ce{nC6H14} COM falls on the central carbon atom. The calculation was carried out only for non-bonded interactions.
 For both organic molecules there are marked oscillations between \SI{0.1}{\nm} and \SI{0.4}{\nm} distances, with considerably deep local minima indicating strong attractions between similar moieties, as expected from their hydrophobic characters. We surmise that the minimum at \SI{0.15}{\nm} corresponds to H--H  stacking of in the cyclohexane \ce{cC6H12} case and to the side-side alignment in the case of n-hexane \ce{nC6H14} (see Figure \ref{fig:characteristic_lenght}) as also suggested by the snapshots of the final equilibrated conformations in Figure \ref{fig:hydrophobicity_snap}.  Also in both cases, the first and the second local minima are located at approximated distances \SI{0.15}{\nm} and \SI{0.25}{\nm}, somewhat closer compared to the results of  \citeauthor{Smith1993} \cite{Smith1993} who performed a similar calculation for two methane molecules and found these two minima located at \SI{0.40}{\nm} and \SI{0.65}{\nm}. The depth of the first deeper minimum ($ \approx $  1 kCal per mole )  is similar to that found for methane in Ref. \cite{Smith1993}.  Several differences are however visible between the \ce{cC6H12}-\ce{cC6H12} interactions (solid magenta line) and the \ce{nC6H14}-\ce{nC6H14} interactions (solid blue line). In the case of \ce{cC6H12}-\ce{cC6H12} interactions (solid magenta line) both the minima and the maxima are rather narrow and deep compared to those of \ce{nC6H14}-\ce{nC6H14} interactions that are much broader (solid blue line). Also the first large and positive energy barrier occurs around  \SI{0.075}{\nm} for \ce{nC6H14}-\ce{nC6H14} interactions (solid blue line) and around  \SI{0.13}{\nm} for \ce{cC6H12}-\ce{cC6H12}. Finally, \ce{cC6H12}-\ce{cC6H12} interactions (solid magenta line) present a third minima which is absent in the \ce{nC6H14}-\ce{nC6H14} counterpart.

These findings are consistent with the idea that the two solvents are expected to interact similarly with water, having very similar chemical properties -- the nearly identical dielectric constants, and very similar geometric sizes (as represented by the diameter of the equivalent van der Waals sphere), but not identically. As cyclohexane \ce{cC6H12} and n-hexane \ce{nC6H14} have different shapes, as displayed in Figure \ref{fig:characteristic_lenght}, this might induce a difference in their solvation entropies that will be discussed further below. 
\subsection{Characterization of the folding process of poly-phenylacetylene (pPA)}
\label{subsec:characterization}
Let us recapitulate our previous findings and put them in the perspective of the next step. Consider the work by \citeauthor{Vasilevskaya2003} \cite{Vasilevskaya2003} who studied the conformational equilibrium properties of amphiphilic polymers in a poor solvent. The amphiphilic polymer was modelled by a chain of hydrophobic monomers (H-type) with a side-chain polar group (P-type), and the poor solvent character was modelled implicitly by modulating the H-H, H-P, and P-P interactions. At constant (room) temperatures, a collapse of the chain was achieved by favouring the H-H and P-P interactions with respect to the H-P ones. This promotes segregation of the H and P groups and hence it induces the collapse of the chain. Now imagine extending this study to \textit{explicit} solvent, as it was done in the bead-spring model. This requires the additional classification of the solvent in hydrophobic (H-type) or polar (P-type). The collapse of the amphiphilic polymer will then be promoted by a P-type (H-type) solvent if monomer character is mainly hydrophobic (polar). In the case of a perfect H-P amphiphilic polymer containing an identical numbers of hydrophobic and polar groups, both H and P-type of solvent will be neutral and the coil-globule transition can occur only upon cooling. 

Poly-phenylacetylene (pPA) (Figure \ref{Fig:initial_linear_20mer}) is a nonbiological polymer that belongs to the general class of aromatic foldamers \cite{Hill2001,Teng2023} that are of considerable interest for its potential technological applications \cite{Freire2016}. A key observation was that pPA chain adopts a nonflat helical structure \cite{Nelson1997,Gin1999,Prince1999,Prince2000,Yang2000,Elmer2001,Sen2002,Elmer2004}, and this opens many perspectives associated with its helicity. 
It is a polymer formed by repeated units of an alkyne hydrocarbon containing a phenyl ring (the monomers) whose rigidity promotes the helical shape upon folding \cite{Qi2012}, whose shape, chirality, and self-assembly properties are the result of a delicate balance between the various interactions, and modification of the solvent polarity and/or the temperature may promote some factors over the others, and hence affect this delicate balance. A deeper understanding of these factors is the main driving force for the present study that builds on important past contributions. \citeauthor{Nelson1997} \cite{Nelson1997} first observed that pPA m-th oligomers (with $m \geq 7$) undergoes a coil-helix transition in a putative poor solvent such as chloroform \ce{CHCl3} (see Figure \ref{fig:characteristic_lenght}). Upon replacing the benzoate side chain \ce{CH3} (Figure \ref{Fig:initial_linear_20mer}) with hydrogen \ce{H}, the folding occurs in water \ce{H2O}, so in this case also water acts as a poor solvent for pPA. The same backbone but with a slightly different side chain was used in a latter study by the same group \cite{Prince1999} and it was found that it remains in a random swollen conformation in a putative good solvent such as acetonitrile \ce{MeCN} (see Figure \ref{fig:characteristic_lenght}). The kinetic of the process has also been studied both experimentally \cite{Yang2000} and numerically \cite{Elmer2001} again for an oligomer different from that considered in the present study (Figure \ref{Fig:initial_linear_20mer}). Two key aspects remaining unclear from these studies are a clear understanding of the driving force for collapse, and a clear definition of "good" and "poor" solvent for pPA. One preliminary step in this direction was performed by \citeauthor{Sen2002} \cite{Sen2002} in the case of \ce{H} side chain, but this study was unable to investigate the full folding process because of it high computational cost out of reach at that time. To the best of our knowledge, the present study is hence the first full-fledged atomistic investigation of this particular oligomer that can be compared with the experiments of Ref. \cite{Nelson1997}. The present study will further offer several additional insights stemming from the use of different solvents.

MD atomistic simulations of a single chain of poly-phenylacetylene (pPA) (with \ce{CH3} side chain, see Figure \ref{Fig:initial_linear_20mer}) with different number $m$ of monomers, from 12 to 20, in solvent of different polarities were performed according to the aforementioned prescription. 
As anticipated, we choose to represent the polarity of a solvent by its (relative) dielectric constant (see Figure \ref{fig:characteristic_lenght}). Hence, solvents ranging from polar (water \ce{H2O}) to hydrophobic (cyclohexane \ce{cC6H12} and \textit{n}-hexane \ce{nC6H14}), were employed. With reference to past work, we note that in this classification, choloroform \ce{CHCl3} (dielectric constant 4.81) is hydrophobic, whereas acetonitrile \ce{MeCN} (dielectric constant 37.5) is polar.  As pPA (with \ce{CH3} side chain) is predominantly hydrophobic in character due to its chemical structure (see Figure \ref{Fig:initial_linear_20mer}), one might expect a random coil conformation in a hydrophobic solvent such as cyclohexane \ce{CC6H12}, and a folded helical conformation in a polar solvent such as water \ce{H2O}. Note, however, that this expectation is already \textit{not} consistent with experimental findings of \citeauthor{Nelson1997} \cite{Nelson1997} who observed the folding of the chain in chloroform \ce{CHCl3} hydrophobic in our classification (dielectric constant 4.81). In all cases, the system was kept at room temperature. As it will be further elaborated below, the temperature may also be expected to play an important role.  For example a polystyrene (predominantly hydrophobic) single chain was observed to collapse in cyclohexane (also hydrophobic) upon cooling from \SI{35.0}{\celsius} to \SI{28}{\celsius} \cite{Chu1995}.
\begin{figure*}[htbp]
\centering
\begin{subfigure}{\textwidth}
\centering
\includegraphics[width=\textwidth, trim=0cm 0cm 0cm 0cm, clip=true]{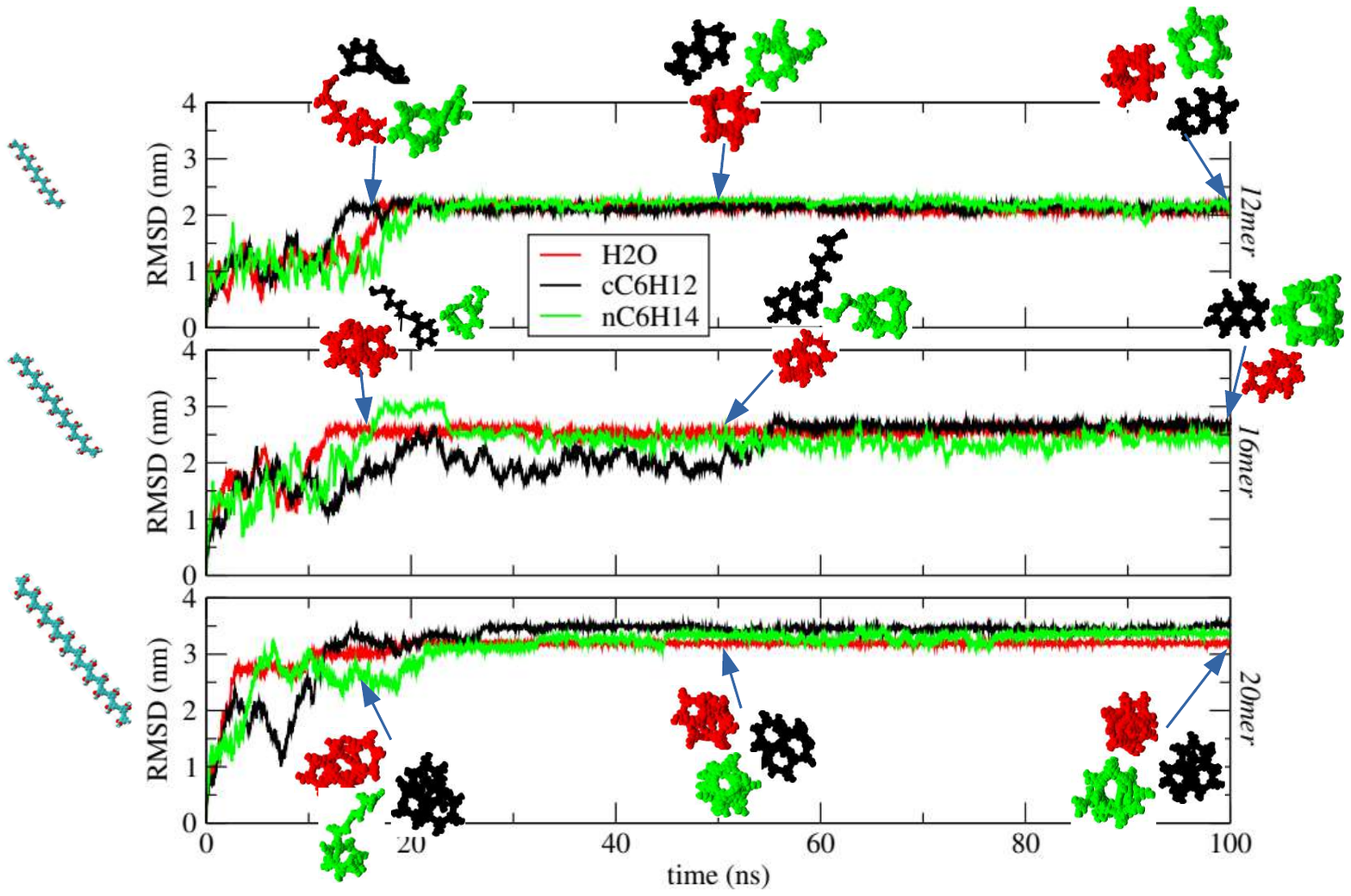}
\end{subfigure}
\caption{Root-Mean-Square-Deviation (RMSD) from the initial conformation along the selected simulation trajectories. All the simulations started from the extended swollen conformations (left most inset view) and representative snapshots are extracted at representative time frames 15, 50, and \SI{100}{\ns}. Simulations performed in water \ce{H2O} (red curve), cyclohexane \ce{cC6H12} (black curve), and n-hexane \ce{nC6H14} (green curve) have also insets displaying representative snapshots with the same color code. From top to bottom results are reported for oligomer lengths \textit{12mer},  \textit{16mer} and \textit{20mer}.}
\label{fig:rmsd_conf_overviews}
\end{figure*}

Our results are reported in Figure \ref{fig:rmsd_conf_overviews}, where the root-mean-square-deviation $\text{RSMD}(t)=\sqrt{\sum[\mathbf{r}_{i}(t)-\mathbf{r}_{i}(0)]^{2}}$ from the initial stretched conformation (depicted in the far left inset of the Figure) is displayed during the time evolution of the system in the case where pPA is dispersed in water \ce{H2O} (red curve), cyclohexane \ce{cC6H12} (black curve) and n-hexane \ce{nC6H14} (green curve). Three different number of monomers ($n=12$, $n=16$, $n=20$) are reported from top to bottom to test for the length dependence. Representative snapshots of the different conformations of the pPA polymer are also displayed in Figure \ref{fig:rmsd_conf_overviews} as insets, color coded according to the relative curve. It is apparent how, for all three solvents and chain lengths, the RSMD displays an initial increase eventually reaching a plateau thus indicating the tendency to form a final globular conformation different from the initial extended conformation. However, while in water the conformational transition appears to be rather stable after a transient period of about \SI{20}{\ns}, both in cyclohexane and (to a less extent) n-hexane the chain appears to fold/unfold erratically, thus suggesting that the folded state is not fully  stable at this temperature in both solvents.
These results also appear to be independent on the length of the polymer, the three panels of Figure \ref{fig:rmsd_conf_overviews} being very similar to one another. In the experiments reported in Refs. \cite{Nelson1997,Prince1999}, acetonitrile \ce{MeCN} (dielectric constant $37.5$) was used as polar solvent, whereas cholorophorm \ce{CHCl3} (dielectric constant $4.81$) was used as nonpolar (hydrophobic) solvent. Our findings appear to be consistent with theirs, as they found a stable fold for octamers or above only in acetonitrile (polar), whereas no folding was observed in cholorophorm (hydrophobic).
\citeauthor{Yang2000} \cite{Yang2000} also reported experimental folding of pPA dodecamer in 50/50 THF/methanol, and \citeauthor{Elmer2004} \cite{Elmer2004} performed detailed numerical simulations and unconvered kinetic trapping occurring during the folding pathway.
\begin{figure*}[htbp]
\centering
\begin{subfigure}[b]{0.490\textwidth}
    \centering
    \includegraphics[width=\textwidth]{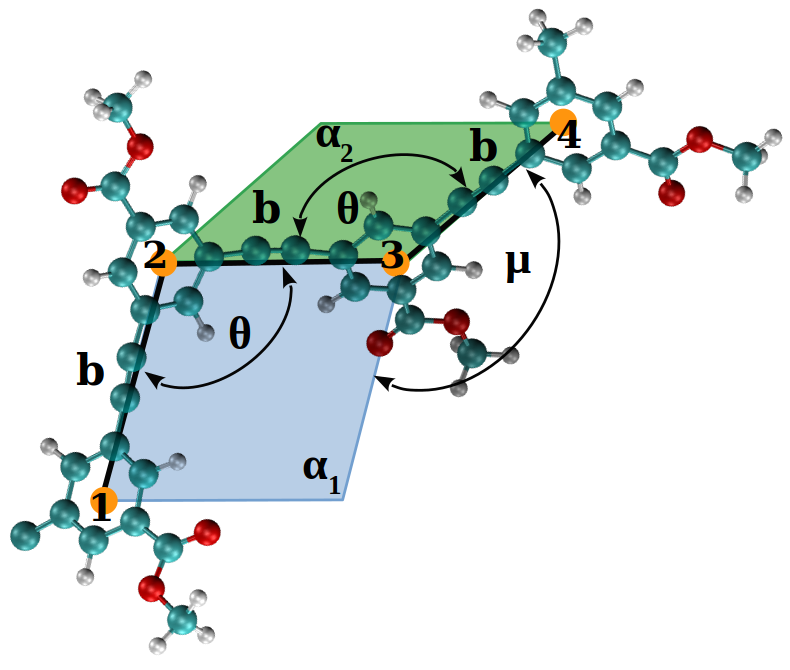}
     \caption{\label{fig:fig8a}}
    \end{subfigure}
    \begin{subfigure}[b]{0.490\textwidth}
    \centering
    \includegraphics[width=\textwidth]{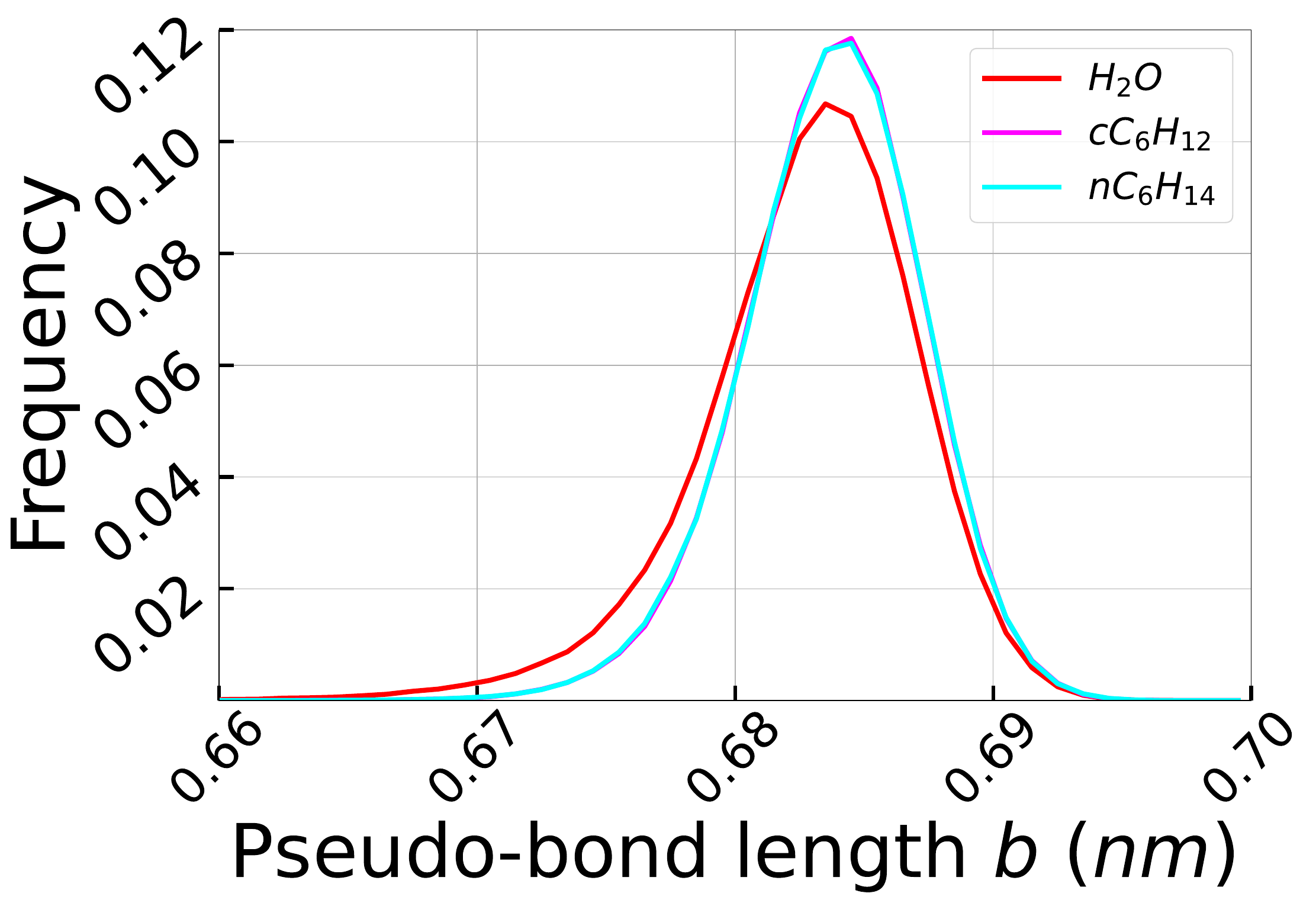}
     \caption{\label{fig:fig8b}}
    \end{subfigure}
    \\
    \begin{subfigure}[b]{0.490\textwidth}
    \centering
    \includegraphics[width=\textwidth]{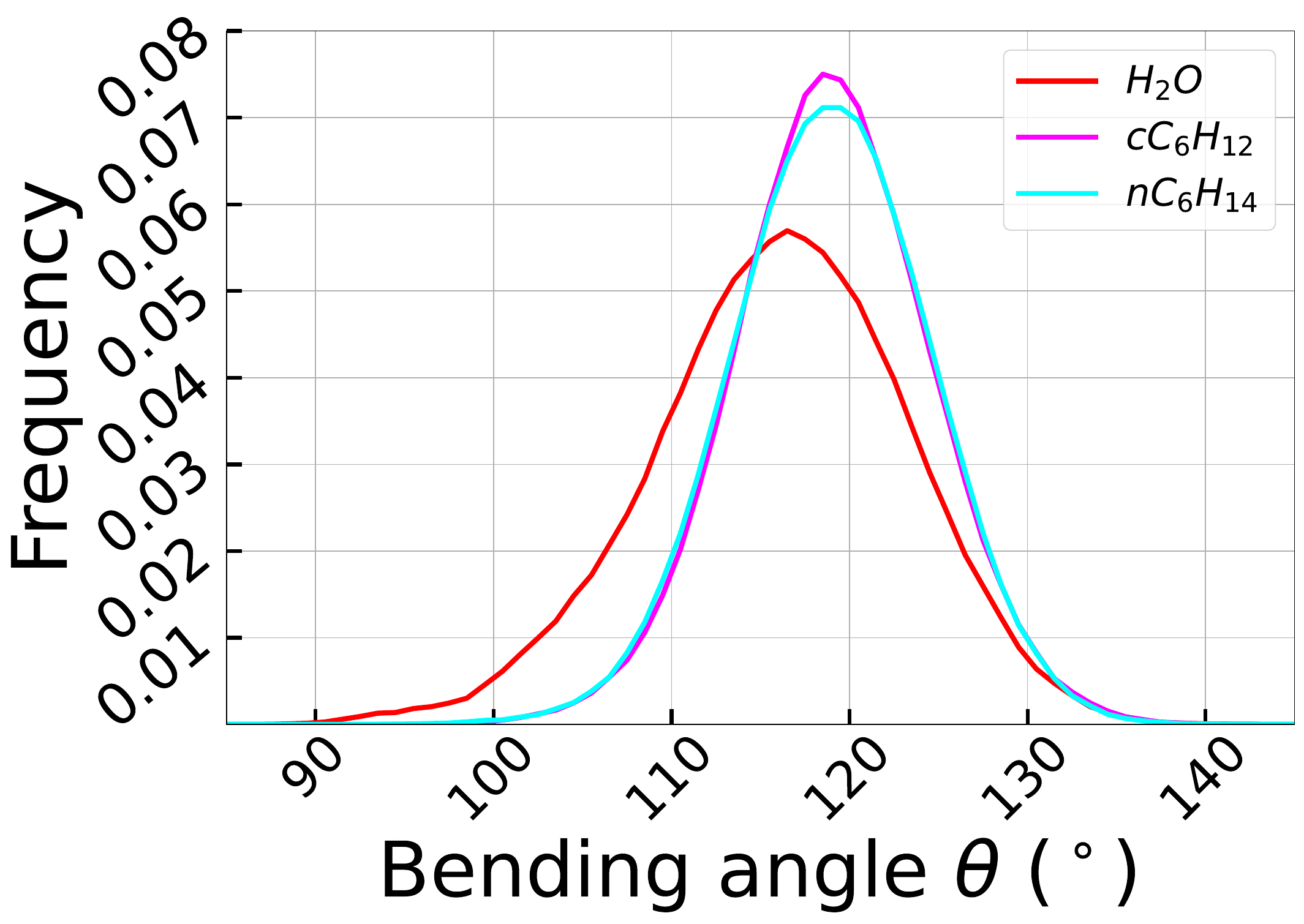}
     \caption{\label{fig:fig8c}}
    \end{subfigure}
    \begin{subfigure}[b]{0.490\textwidth}
    \centering
    \includegraphics[width=\textwidth]{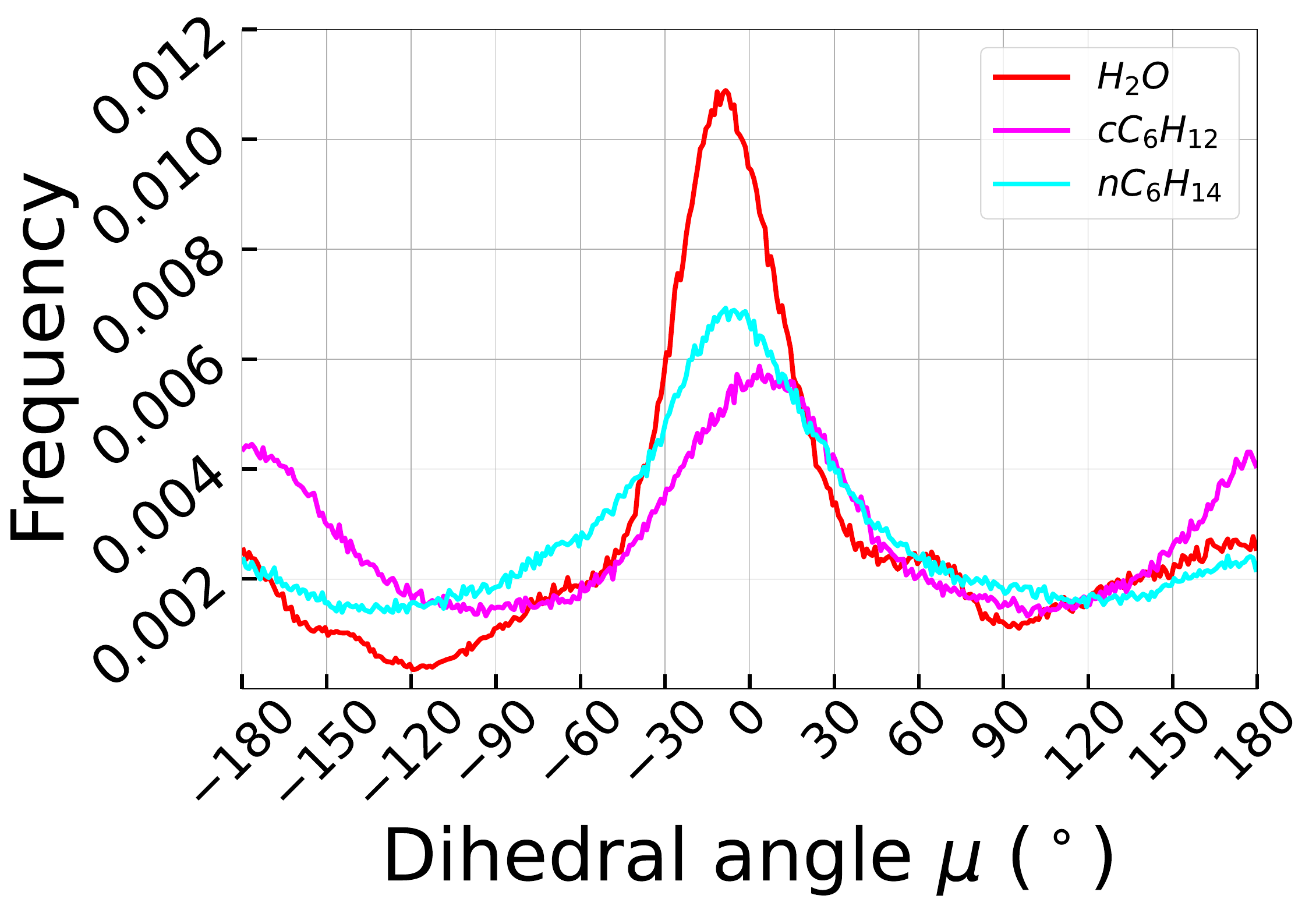}
     \caption{\label{fig:fig8d}}
    \end{subfigure}
  \caption{(a) Definitions of the pseudo-bond (Kuhn) length $b$, bond bending angle $\theta$, and dihedral angle $\mu$ \cite{Skrbic2021}; (b) Distribution of the pseudo-bond length $b$; (c) Distribution of bond bending angle $\theta$; (d) Distribution of the dihedral angle $\mu$ . Results include different oligomer units \textit{12mer}, \textit{16mer} and \textit{20mer}, and different curves refer to the three different solvents, water \ce{H2O} (red), cyclohexane \ce{cC6H12} (magenta) and \textit{n}-hexane \ce{nC6H14} (cyan) employed in this work.} 
  \label{fig:mm_dist}
\end{figure*}
A glance to the most representative snapshots reported in Supplementary Material shows how within each helical structure (either stable or unstable), the highly charged oxygen atom of each monomer tends to avoid a close contact with the oxygen atoms of the non-bonded neighboring monomers to minimize the high repulsive energy, although this is partially mitigated by the presence of the hydrogen atoms within the same monomers. This was noted before \cite{Srikanta2002}. Although this prevents the achievement of a perfect $\pi-\pi$ stacking between consecutive rings, the next analysis shows that in water this turns out not to be a crucial feature.
Additional insights can be obtained by accumulating the pseudo-bond length $b$ (i.e. the effective distance between monomers, also known as Kuhn length \cite{Khokhlov2002}), the bending angle $\theta$ between three consecutive  $i-1$,$i$, and $i+1$ monomers, and the dihedral angle $\mu$, the angle between two consecutive planes involving the $i-1$,$i$, $i+1$, and $i+2$ monomers as shown in Figure \ref{fig:fig8a} \cite{Skrbic2021}, and whose distributions along the equilibrated trajectory is reported in Figures \ref{fig:fig8b}- \ref{fig:fig8d}. Here, the representative point of the monomer is the geometrical center of its phenyl ring, see the top panel of Figure \ref{Fig:initial_linear_20mer}. Coherently with previous findings, the distribution of the pseudo-bond lengths in water \ce{H2O} is peaked around 6.85 \AA, with that in both cyclohexane \ce{cC6H12} and n-hexane \ce{nC6H14}  both peaked at a slightly larger value (Figure \ref{fig:fig8a}). A more significant difference is observed for the bending angle $\theta$ whose distribution is found to centered at $\theta \approx 115^{\circ}$ in water and at $\theta \approx 118^{\circ}$ in both cyclohexane and n-hexane (Figure \ref{fig:fig8c}). An even more marked difference is finally observed for the distribution of the dihedral angles $\mu$ whose distribution is found to be centered at $\mu \approx -10^{\circ}$ in water, indicating a stable $\pi-\pi$ stacking of the consecutive phenyl rings thus eventually leading to the helical structure ( red curve Figure \ref{fig:fig8d} bottom panel). By contrast, both cyclohexane and n-hexane display additional peaks in the dihedral angle $\mu$ close to $\mu \approx \pm 180^{\circ}$ indicating that the fold/unfold dynamical process occurs with a high frequency even after the equilibration has been achieved, in agreement with the results of Figure \ref{fig:rmsd_conf_overviews}. As in the study by \citeauthor{Nelson1997} \cite{Nelson1997}, in all cases the collapsed structures were found to be helices as reported in Supplementary Material. As mentioned, this is to be ascribed to the rigidity of the aromatic chain forming the chain backbone, which forces the chain to fold into a helicoidal structure via $\pi-\pi$ stacking, and it is fully formed only when the chain length (that is the number of monomers) is sufficiently long and commensurate with the pitch of the chain.

\textcolor{black}{Additional insights can be obtained by computing the intramolecular site-site pair radial distrubution function between carbon atoms of pPA methoxycarbonyl group in the three solvents. This is reported in Supplementary Material and displays clear peaks approximately 0.5nm apart in water and n-hexane, a value that matches rather well with the distance between the aromatic planes in the $\pi-\pi$ stacking \cite{Spano2023}. By contrast, in cyclohexane, the peak distribution is fuzzier coherently with an unstable fold.}

Within the common paradigm of good and poor solvent outlined in the paradigmatic case of a bead-spring polymer in a Lennard-Jones fluid, the above findings can be interpreted in terms of a polymer backbone mainly hydrophobic that collapses in water environment but not in an organic hydrophobic solvent such as cyclohexane or n-hexane. This is also consistent with another study \cite{Wijesinghe2016} where MD simulations of a single carboxylate-substituted poly \textit{para phenylene ethynylene} (PPE) chain  in different solvents (water \ce{H2O} and toluene \ce{C6H5CH3}) were performed. In that case too, the chain was found to remain extended at room temperature in toluene, and found to collapse in water.

The results of Fig.\ref{fig:rmsd_conf_overviews} appears to have only a minimal dependence on the length of the polymer, with nearly identical behaviour in the case of oligomer lengths \textit{12mer},  \textit{16mer} and \textit{20mer}. This was to be expected.
 \citeauthor{Nelson1997} \cite{Nelson1997} observed  experimental folding of a phenylacetylene oligomer only for oligomer length 6 or above, with the rigid nature of the phenyl ring preventing $\pi-\pi$ stacking for too short chains. Likewise \citeauthor{Prince1999} \cite{Prince1999} observed the collapse of the chain in acetonitrile (polar) for  oligomer length 8 or above. As all oligomer lengths reported in Figure \ref{fig:rmsd_conf_overviews} are in all cases sufficiently long to allow $\pi-\pi$ stacking, it is not surprising that the results turn out to be independent of the lengths.
\subsection{Additional quantitative probes of the folding process}
\label{subsec:quantitative}
\begin{figure}[htbp]
\centering
\begin{tikzpicture}
\node[anchor=south west,inner sep=0] (image) at (0,0){\scalebox {0.3}{\includegraphics[width=1.6\textwidth, trim=0cm 0cm 0cm 0cm, clip=true, angle=0, page=1]{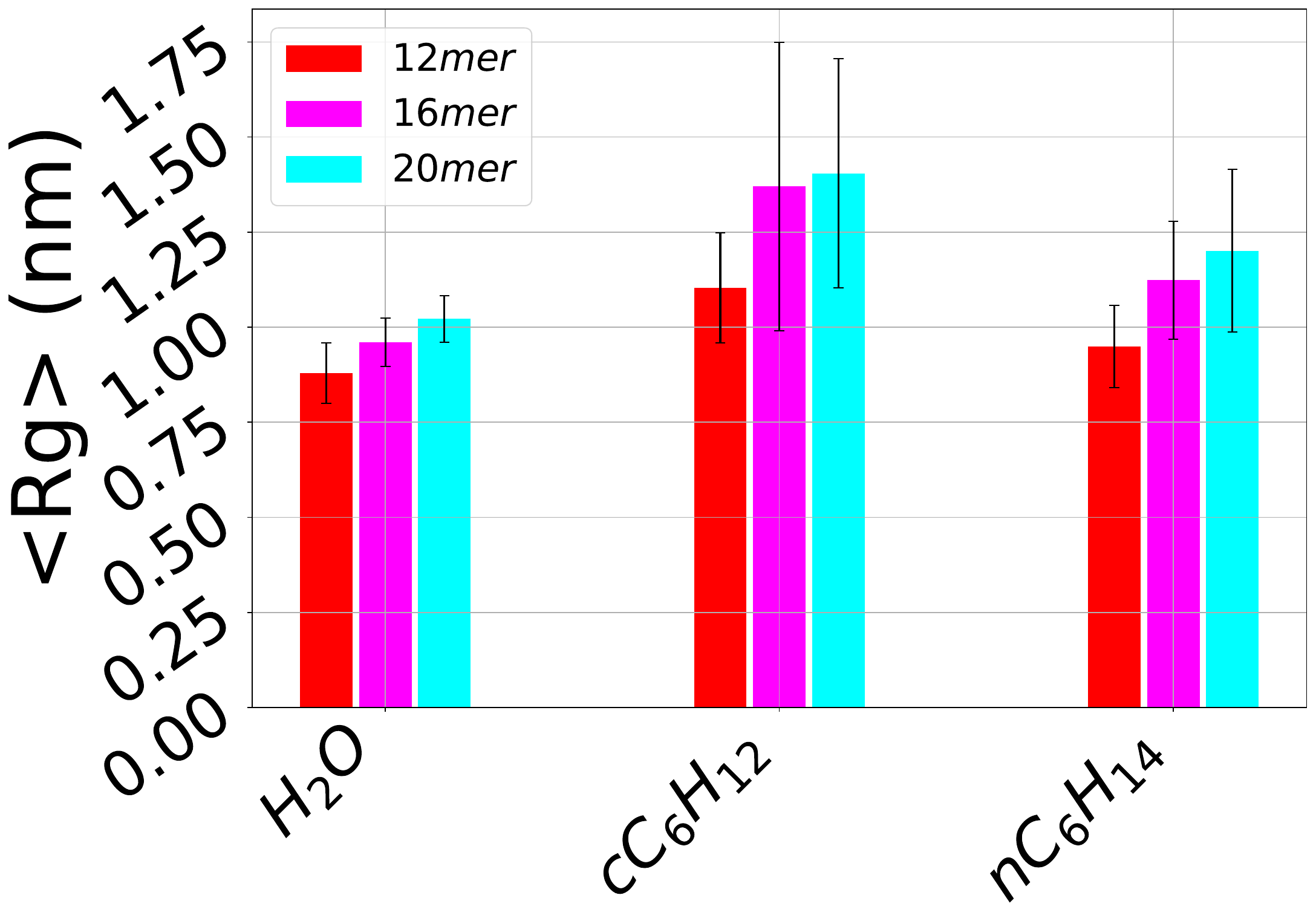}}};
\end{tikzpicture}
\begin{tikzpicture}
\node[anchor=south west,inner sep=0] (image) at (0,0){\scalebox {0.3}{\includegraphics[width=1.6\textwidth, trim=0cm 0cm 0cm 0cm, clip=true, angle=0, page=1]{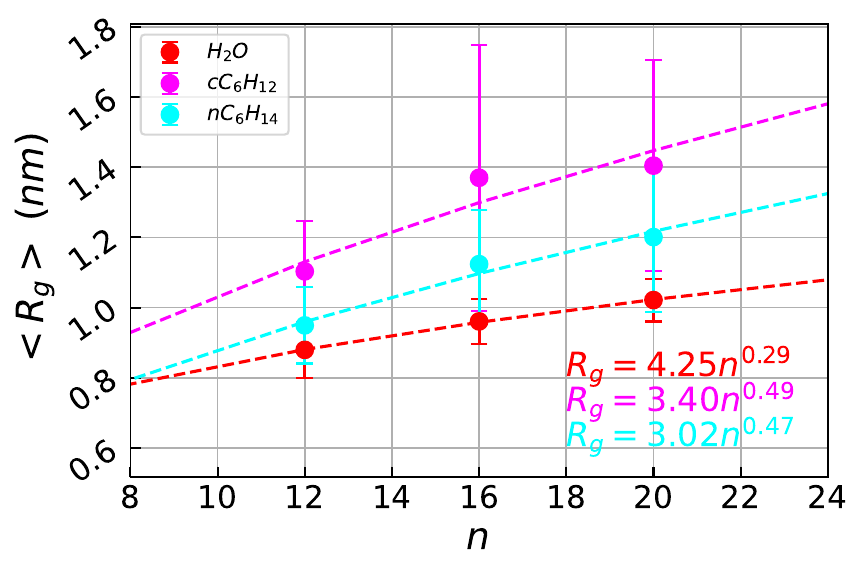}}};
\end{tikzpicture}
\caption{(Left) Average \textit{pPA} radius of gyration $\langle R_g \rangle$ for the individual runs performed. The error bars represent the standard deviations. All the three solvents water \ce{H2O}, cyclohexane \ce{cC6H12}, and \textit{n}-hexane \ce{nC6H14} are displayed for each of the oligomer's length \textit{12mer}, \textit{16mer} and \textit{20mer} considered here. The related time-based plots are shown in Supplementary Material. (Right) Average \textit{pPA} radius of gyration versus the number of monomers $n$ for the three different solvents water \ce{H2O}, cyclohexane \ce{cC6H12}, n-hexane \ce{nC6H14}. Fitted slopes are also reported in the three cases. These results were obtained at room temperature of \SI{300}{\kelvin}.}
\label{fig:rg_avg}
\end{figure}

To provide additional evidence of the folding process, we also monitored the the radius of gyration $R_g=\sqrt{\sum_{i} m_i[\mathbf{r}_{i}-\mathbf{R}_{CM}]^{2}/M}$ ($m_i$ is the mass of the $i$-th atom, $\mathbf{R}_{CM}$ is the center of mass of the polymer chain, and $M$ is the total mass), and the solvent accessible surface area (SASA) using the algorithm devised by \cite{Eisenhaber1995}. We note that SASA can be regarded as a proxy of the sum of the cavity and van der Waals contributions to the solvation free energy \cite{Leach2001} and hence its monitoring is particularly insightful. For five independent runs, the average $\langle R_g \rangle$ and the standard deviation are reported in Figure \ref{fig:rg_avg} (left). Three different number of monomers $n=12,16,20$ were considered for all three different solvents.  The same quantity is also plotted as a function of the number of monomers $n$ in Figure \ref{fig:rg_avg} (right), to identify the various phases. We find that $\langle R_g \rangle \approx n^{\nu}$ with $\nu \approx 0.29$ for water \ce{H2O}, in reasonable agreement with the value $\nu=1/3$ expected for the collapsed phase; $\nu \approx 0.49$ in the case of cyclohexane \ce{cC6H12}, and  $\nu \approx 0.47$ in the case of n-hexane \ce{nC6H14}, both somewhat lower of the Flory value $\nu=3/5$ expected for the swollen phase in a good solvent \cite{DeGennes1979,Doi1988}, likely indicative of the aforementioned erratic folding/unfolding process. 
It is here worth recalling that, strictly speaking, the above scaling laws are only valid in the thermodynamic limit, that is for sufficiently long polymers \cite{DeGennes1979}, and hence a deviation from these laws is to be expected for short oligomers such as those discussed here.

\begin{figure*}[htpb]
\centering
\begin{subfigure}[b]{0.40\textwidth}
\includegraphics[width=\textwidth]{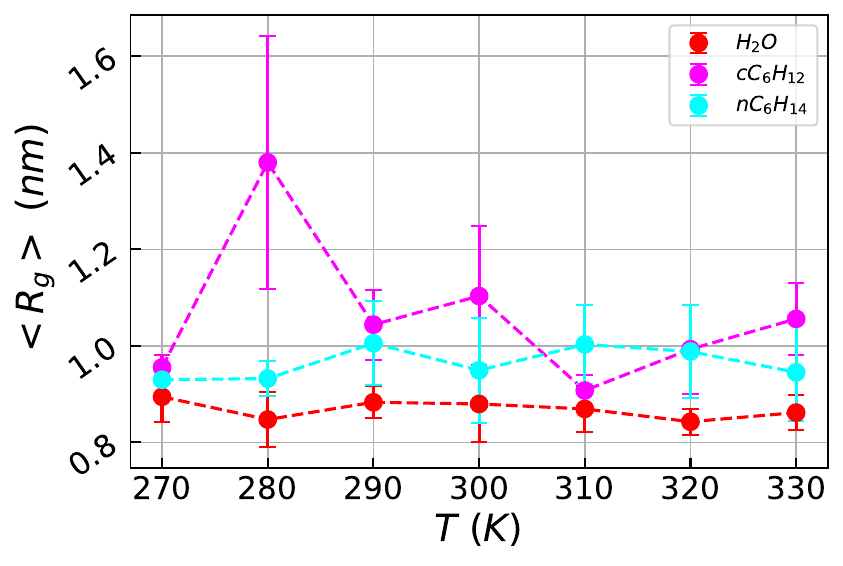}
\caption{12mer}
\label{fig:rg_avg_T_12mer}
\end{subfigure}
\begin{subfigure}[b]{0.40\textwidth}
\includegraphics[width=\textwidth]{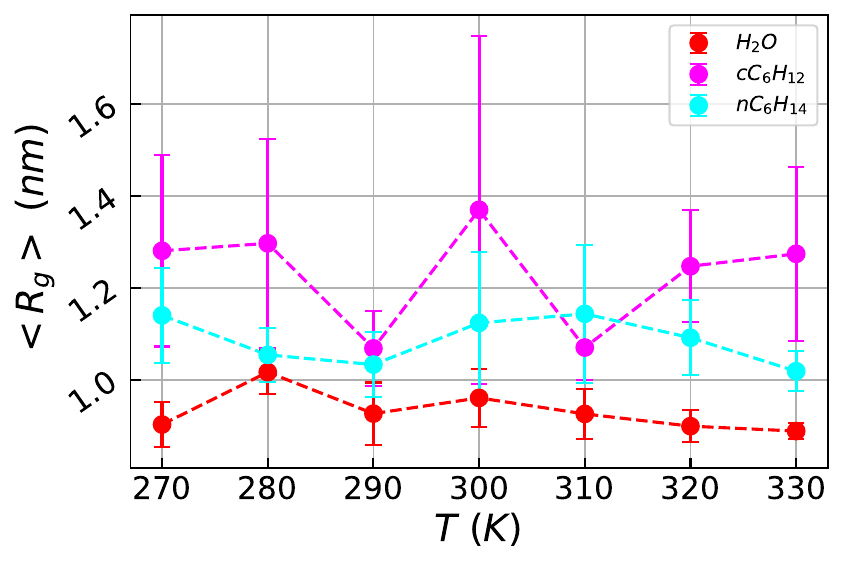}
\caption{16mer}
\label{fig:rg_avg_T_16mer}
\end{subfigure} \\
\begin{subfigure}[b]{0.40\textwidth}
\includegraphics[width=\textwidth]{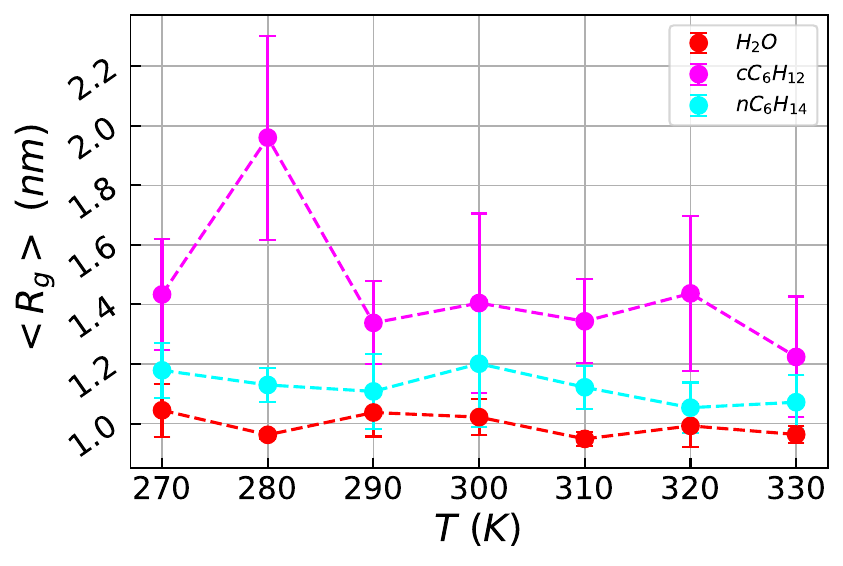}
\caption{20mer}
\label{fig:rg_avg_T_20mer}
\end{subfigure}
\begin{subfigure}[b]{0.40\textwidth}
\includegraphics[width=\textwidth]{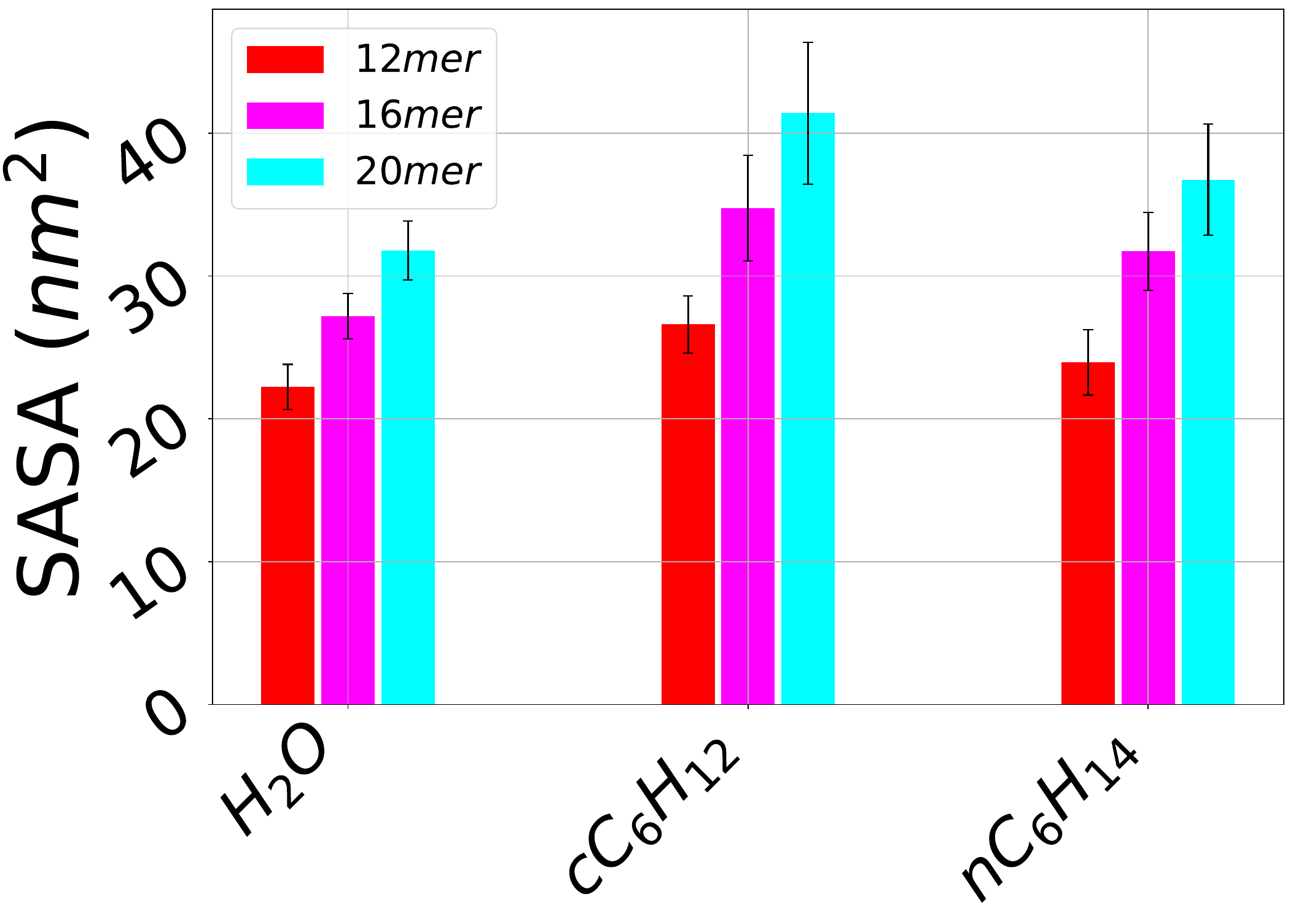}
\caption{SASA}
\label{fig:sasa_avg}
\end{subfigure}
\caption{Temperature dependence of the average \textit{pPA} radius of gyration for different oligomer length (a) \textit{12mer}, (b) \textit{16mer} and (c) \textit{20mer} in the three solvents water \ce{H2O}, cyclohexane \ce{cC6H12}, and \textit{n}-hexane \ce{nC6H14} considered here. (d) Average \textit{pPA} accessible surface area over the independent simulations performed at \SI{300}{\kelvin} temperature. The error bars stand for standard deviations. All the three solvents water \ce{H2O}, cyclohexane \ce{cC6H12}, and \textit{n}-hexane \ce{nC6H14} are displayed along with the oligomer's length \textit{12mer}, \textit{16mer} and \textit{20mer} used in this work. The corresponding time-based plots are shown in Supplementary Material.  }
\label{fig:rg_avg_T}
\end{figure*}

The results of Figure \ref{fig:rg_avg} were obtained at room temperature of \SI{298.15}{\kelvin}. However, the folding process appears rather stable against a temperature change in the interval \SI{270}{\kelvin}- \SI{330}{\kelvin}, which brackets the \SI{298.15}{\kelvin} temperature and coincides with thar previously used for oligopeptides \cite{Dongmo2023}. This is shown in Figure \ref{fig:rg_avg_T} where the average radius of gyration $\langle R_g \rangle$ is reported at a function of the temperature $T$ in the temperature range \SI{270}{\kelvin}- \SI{330}{\kelvin}. Although we eventually expect to have a marked collapse of the chain for sufficiently low temperatures for all three solvents, the apparent insensitivity in this temperature range is rather intriguing. 

The analysis of SASA  at \SI{298.15}{\kelvin} temperature provides consistent results. This is reported in Figure \ref{fig:sasa_avg} where SASA values are displayed for same three different number of monomers $n$ and for the three different solvents: water \ce{H2O}, cyclohexane \ce{cC6H12}, and n-hexane \ce{nC6H14}. For all different $n$, SASA in water shows consistently lower values compared to both cyclohexane and n-hexane,  indicative of a more stable collapsed phase, in agreement with the results of Figures \ref{fig:rmsd_conf_overviews} and \ref{fig:rg_avg}.

\textcolor{black}{A final conclusion can be drawn for the occurring diffusion process of the folded structures. As the collapse (when present) is from a coil to a helical-like structure, rather than a globule, the diffusion need not be isotropic. Indeed, by recording the mean square displacements (MSD) with respect to the original conformation $\langle \big( \mathbf{R}_{GC}(t) - \mathbf{R}_{GC}(0) \big)^2 \rangle = 6D t$ of the geometrical centers of 20 aromatic rings mid-points (GC) of the pPA polymer (in units of $10^{3}nm{}^2$) \textit{vs} time (in $ns$), along with the corresponding parallel  $\langle \big( \mathbf{R}_{GC,||}(t) - \mathbf{R}_{GC,||}(0) \big)^2 \rangle = 2D_{||} t$, along the helical axis direction, and perpendicular $ \langle \big( \mathbf{R}_{GC,{\perp}}(t) - \mathbf{R}_{GC,{\perp}}(0) \big)^2 \rangle = 4D_{{\perp}} t$ counterparts, we note that diffusion is systematically faster along the direction parallel to helical axis with respect to the direction perpendicular to it, that is $D_{\parallel} > D_{\perp} > D$, as it is also the case for diffusion of rigid helices \cite{DalCompare2023}. Details can be found in Supplementary Material.}

\subsection{Free energy landscape}
\label{subsec:free_energy_landscape}

\begin{figure*}[htbp]
\centering
\begin{tikzpicture}
\node[anchor=south west,inner sep=0] (image) at (0,0){\scalebox {0.30}{\includegraphics[width=\textwidth, trim=0cm 0cm 0cm 0cm, clip=true, angle=0, page=1]{Fig11a.pdf}}};
\end{tikzpicture}
\begin{tikzpicture}
\node[anchor=south west,inner sep=0] (image) at (0,0){\scalebox {0.30}{\includegraphics[width=\textwidth, trim=0cm 0cm 0cm 0cm, clip=true, angle=0, page=1]{Fig11b.pdf}}};
\end{tikzpicture}
\begin{tikzpicture}
\node[anchor=south west,inner sep=0] (image) at (0,0){\scalebox {0.30}{\includegraphics[width=\textwidth, trim=0cm 0cm 0cm 0cm, clip=true, angle=0, page=1]{Fig11c.pdf}}};
\end{tikzpicture}\\
\vskip-0.12cm
\begin{tikzpicture}
\node[anchor=south west,inner sep=0] (image) at (0,0){\scalebox {0.30}{\includegraphics[width=\textwidth, trim=0cm 0cm 0cm 0cm, clip=true, angle=0, page=1]{Fig11d.pdf}}};
\end{tikzpicture}
\begin{tikzpicture}
\node[anchor=south west,inner sep=0] (image) at (0,0){\scalebox {0.30}{\includegraphics[width=\textwidth, trim=0cm 0cm 0cm 0cm, clip=true, angle=0, page=1]{Fig11e.pdf}}};
\end{tikzpicture}
\begin{tikzpicture}
\node[anchor=south west,inner sep=0] (image) at (0,0){\scalebox {0.30}{\includegraphics[width=\textwidth, trim=0cm 0cm 0cm 0cm, clip=true, angle=0, page=1]{Fig11f.pdf}}};
\end{tikzpicture}\\
\vskip-0.12cm
\begin{tikzpicture}
\node[anchor=south west,inner sep=0] (image) at (0,0){\scalebox {0.30}{\includegraphics[width=\textwidth, trim=0cm 0cm 0cm 0cm, clip=true, angle=0, page=1]{Fig11g.pdf}}};
\end{tikzpicture}
\begin{tikzpicture}
\node[anchor=south west,inner sep=0] (image) at (0,0){\scalebox {0.30}{\includegraphics[width=\textwidth, trim=0cm 0cm 0cm 0cm, clip=true, angle=0, page=1]{Fig11h.pdf}}};
\end{tikzpicture}
\begin{tikzpicture}
\node[anchor=south west,inner sep=0] (image) at (0,0){\scalebox {0.30}{\includegraphics[width=\textwidth, trim=0cm 0cm 0cm 0cm, clip=true, angle=0, page=1]{Fig11i.pdf}}};
\end{tikzpicture}
\caption{Free Energy Landscape (FEL) for each of the oligomers used here and in different solvents, using the average radius of gyration $\langle R_g \rangle$ and the RMSD as reaction coordinates. From top to bottom the FEL in water \ce{H2O}, cyclohexane \ce{cC6H12} and \textit{n}-hexane \ce{nC6H14} are reported for increasing number of monomers from left to right  (\textit{12mer}, \textit{16mer} and \textit{20mer} ). The remaining FEL for all the simulations performed are reported in Supplementary Material. Color coding for the depth of the minima is reported on the right vertical bar. }
\label{fig:fel}
\end{figure*}

The analysis reported in previous section clearly shows how the RMSD, the average radius of gyration $\langle R_g \rangle$, and SASA can all be used as possible "reaction coordinates" to track down and assess the folding/unfolding process. Accordingly, we can construct the relative energy landscape by monitoring their joint probability distribution, and then the relative free energy landscape by using Eq.(\ref{pmf}). Following a common choice in the literature, we selected $\langle R_g \rangle$ and the RMSD as reaction coordinates and compute the free energy landscape  for all the three solvents and different number of monomers. Different combinations involving also SASA provide essentially identical results (see Supplementary Materials).

Figure \ref{fig:fel} reports the results of this free energy landscape analysis, with the first row corresponding to water \ce{H2O}, the second row to cyclohexane \ce{cC6H12}, and the third row to n-hexane \ce{nC6H14}. For each row, different columns refer to different lengths of the polymer, $n=12$ on the left, $n=16$ in the center, and $n=20$ on the right. 
The color code of the contour plots is reported on the right vertical bar and it goes from a shallow minimum (red) to a deep one (blue). 

In water \ce{H2O} (first row in Figure \ref{fig:fel} for $n=12$, $n=16$, and $n=20$ from left to right), a development of a well defined and stable minimum occurring at low $\langle R_g \rangle$ and high RMSD is observed upon increasing the polymer length $n$, clearly associated with a folded conformation. A less pronounced effect is seen in the case of n-hexane \ce{nC6H14} (third row in Figure \ref{fig:fel} for $n=12$, $n=16$, and $n=20$ from left to right), with the depth of the minima much less marked.  By contrast, in ciclohexane \ce{cC6H12} (second row in Figure \ref{fig:fel} for $n=12$, $n=16$, and $n=20$ from left to right) we see evidence of two marginally stable minima, one corresponding to the collapsed phase as in water and the other to the extended phase, indicating the presence of frequent folding/unfolding events, in agreement with the results reported in previous Sections.

Taken together, the above results indicate that at room temperature a pPA oligomer tend to have a stable collapse water \ce{H2O}, an unstable conformation with erratic folding/unfolding events in cyclohexane \ce{cC6H12}, and something in between for n-hexane \ce{nC6H14}, confirming the interpretation of past experimental findings \cite{Nelson1997,Prince1999}. 

In polypeptides, the folded state in water is stabilized also by the formation of intrachain hydrogen bonds \cite{Pauling1951}. In the case of pPA, this is not the case as discussed in the next Subsection.
\subsection{Intra- and inter- oligomer-water hydrogen bonds}
\label{subsec:intra}
\begin{figure*}[htbp]
\centering
\begin{tikzpicture}
\node[anchor=south west,inner sep=0] (image) at (0,0){\scalebox {0.85}{\includegraphics[width=0.7\textwidth, trim=0cm 0cm 0cm 0cm, clip=true, angle=0, page=1]{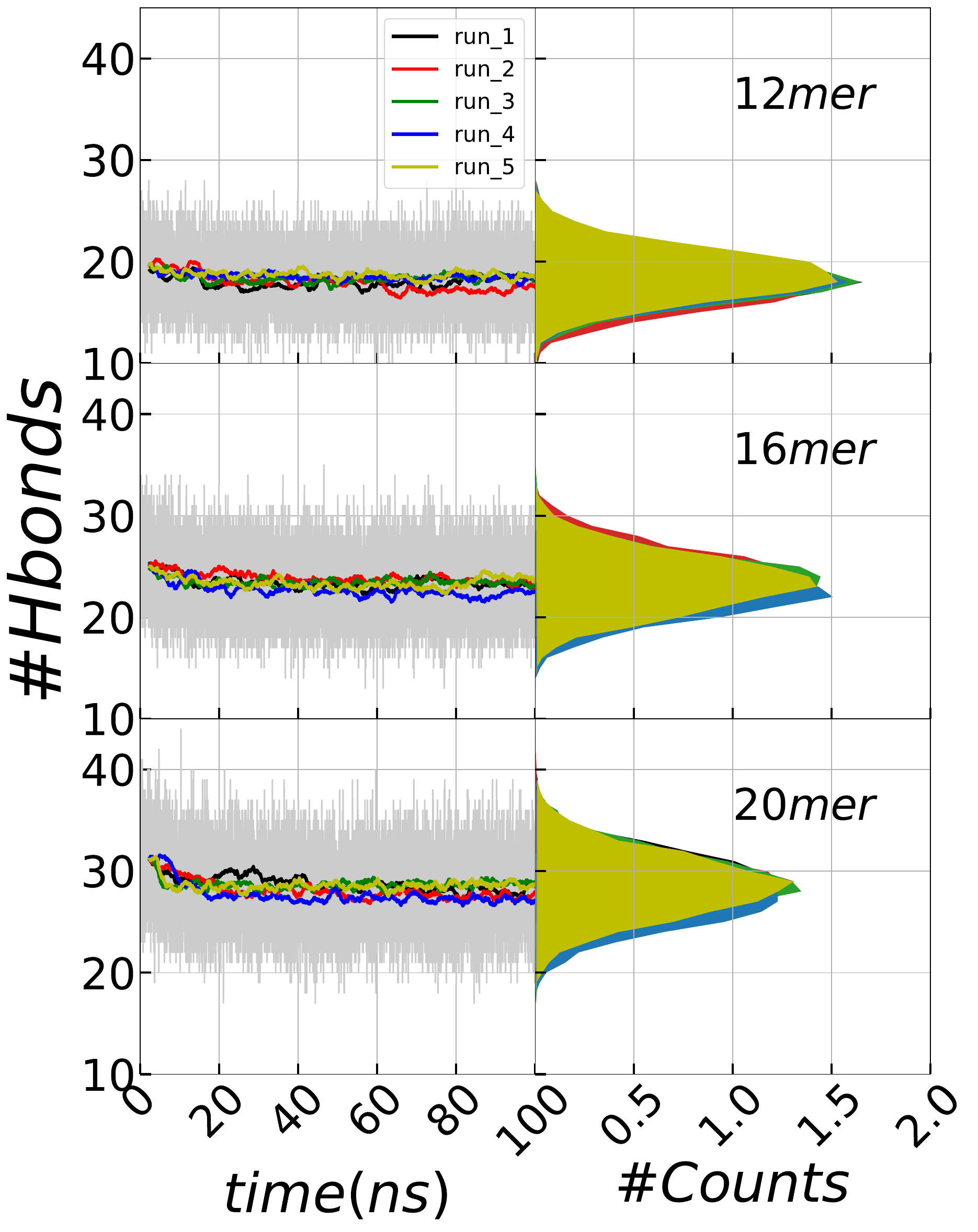}}};
\end{tikzpicture}
\caption{(Left panel) Overview in the changes of the number of solute-solvent hydrogen bonds as a function of the simulation time in water \ce{H2O} for 5 independent runs. From top to bottom the results refer to \textit{12mer}, \textit{16mer} and \textit{20mer}  respectively. The gray fluctuations represent the overlays of the simulation results while their corresponding running average are plotted in the middle. (Right panel) Distribution of hydrogen-bonds during the trajectory. All the plots are on the same scale. 
\label{fig:hbond_change}}
\end{figure*}

The above findings can only partially be interpreted within the common view of "like-dissolves-like". This limitation was noted before in surfactants with inverted polarities \cite{Carrer2020} and in the case of oligopeptides \cite{Dongmo2023}.
One additional interesting question is related to the hydrogen bonds that pPA forms in water \ce{H2O} -- pPA obviously does not form hydrogen bonds with either cyclohexane \ce{cC6H12} or n-hexane \ce{nC6H14}. In the folding process of polypeptides in water \cite{Dongmo2023}, some hydrogen bonds that the peptide chain in the original swollen conformation forms with water break down upon folding and then reform internally as a intra-chain hydrogen bonds. This is indeed a way to stabilize the characteristic  secondary structure in proteins as discovered by Linus Pauling approximately 70 years ago \cite{Pauling1951}. As we will see below, this is \textit{not} the case in the case of pPA.

Figure \ref{fig:hbond_change} (left panel) reports the change in the number of pPA-water hydrogen bonds during the entire MD trajectory in water \ce{H2O}. The three different panels (top, middle, bottom) refers to different lengths of the pPA oligomers (\textit{12mer}, \textit{16mer} and \textit{20mer}), and in all cases, the right panels depict the distribution of the hydrogen bonds. Five independent runs were performed and the colored curve displayed in Figure \ref{fig:hbond_change} report the running average, whereas the underlying gray region represents the relative thermal fluctuations. 
Irrespective of the length of the pPA oligomer, these results unambiguously show that the total number of pPA-water hydrogen bonds does not change significantly during the folding of the pPA oligomer. Hence the optimization of the hydrogen bond distribution is \textit{not} one of the driving force to fold, unlike what happens in the biopolymer case. This suggests that the main driving force for folding in water should be sought elsewhere, and this will be discussed in the next Subsection. 
\subsection{Non-covalent interactions}
\label{subsec:non_covalent}
\begin{figure}[htbp]
\centering
\begin{subfigure}[b]{0.45\textwidth}
\centering
\includegraphics[width=\textwidth]{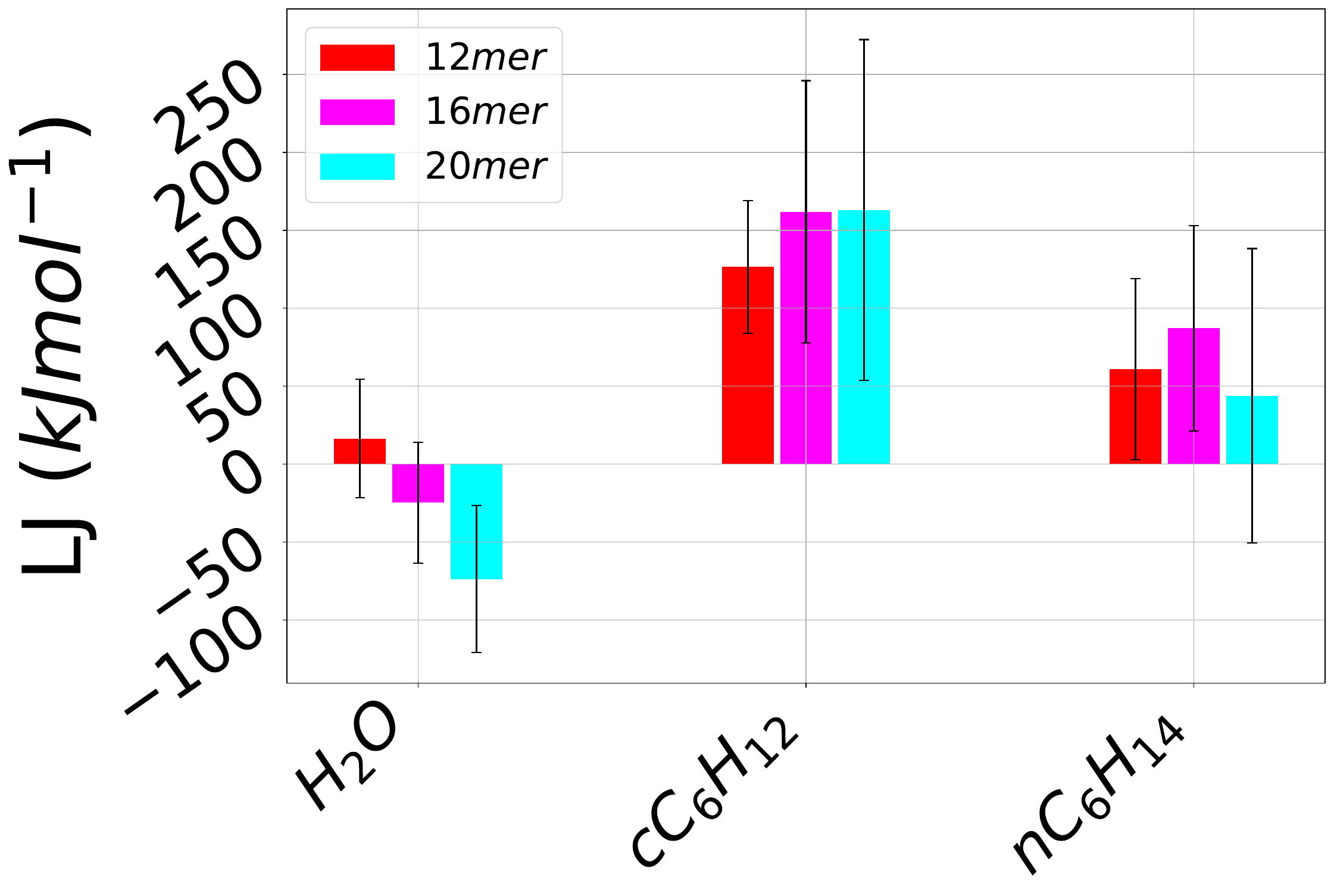}
\caption{\label{fig:fig13a}}
\end{subfigure}
\begin{subfigure}[b]{0.45\textwidth}
\centering
\includegraphics[width=\textwidth]{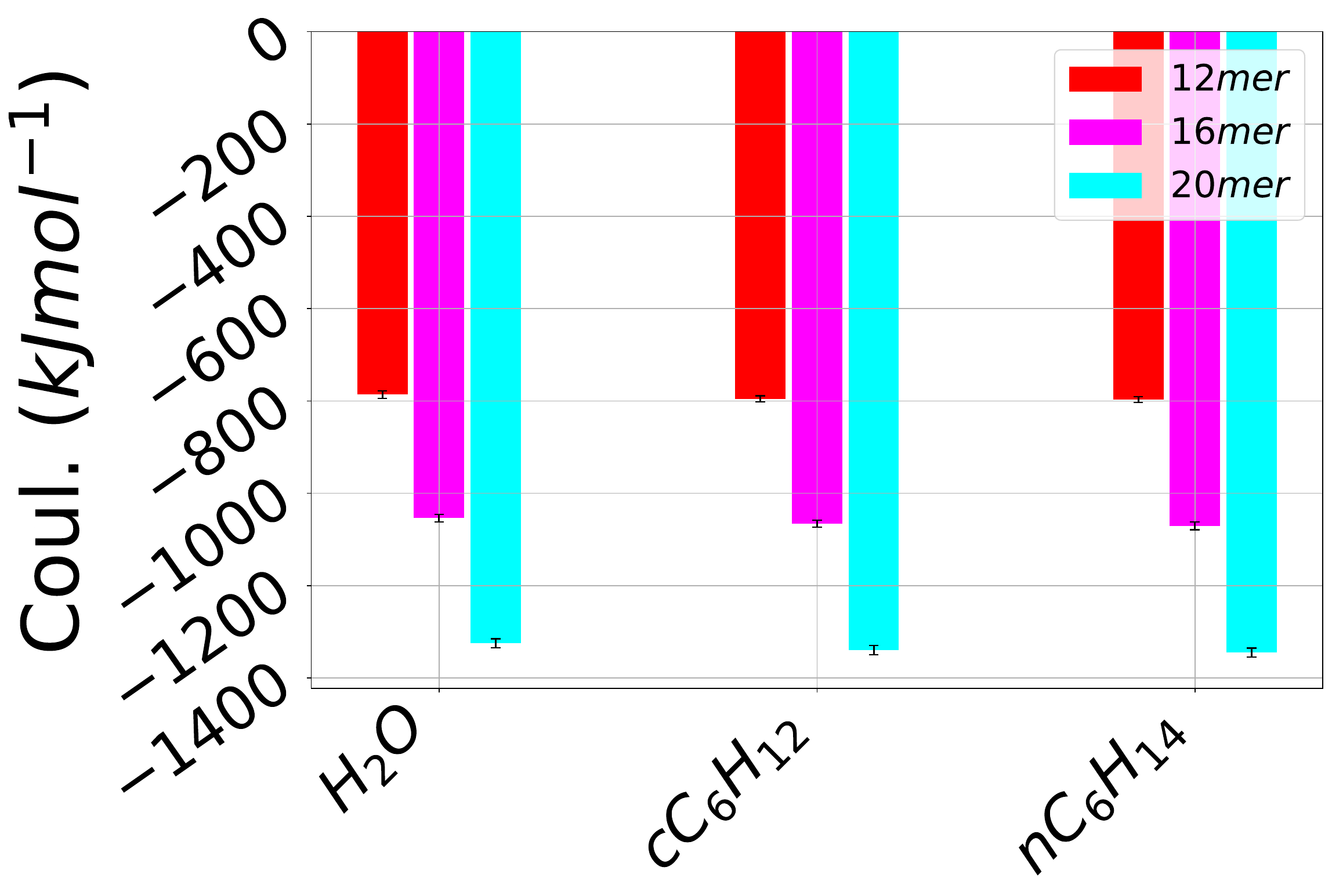}
\caption{\label{fig:fig13b}}
\end{subfigure}
\caption{Average \textit{pPA} intramolecular non-covalent interaction energy over the independent simulations carried out here. (a) Lennard-Jones 12-6 interactions; (b) Coulomb interactions. The error bars stand for standard deviations. All the three solvents water \ce{H2O}, cyclohexane \ce{cC6H12}, and \textit{n}-hexane \ce{nC6H14} are displayed for each of the oligomer's length \textit{12mer}, \textit{16mer} and \textit{20mer}. The reference time-based plots are shown in Supplementary Material.} 
  \label{fig:avg_non_covalent}
\end{figure}
As discussed in the previous Subsection, hydrogen bonds are not a stabilizing factor of the folded state and hence the question arises of what are the other potential stabilizing factors. Hydrogen bonding is present only in water but as it does not change upon folding, its contribution in the total energetic balance can be neglected altogether.
Then, we next discuss the additional energetic part of the non-covalent interactions, Lennard-Jones and Coulomb interactions, that are present for all three considered solvents.
 Figure \ref{fig:avg_non_covalent} depicts the contribution of the Lennard-Jones interactions (Figure \ref{fig:fig13a}) contrasted with the Coulomb contribution (Figure \ref{fig:fig13b}). In both cases, different oligomer lengths have been considered.
In the case of water \ce{H2O}, the Lennard-Jones contribution is negative, decreasing for increasing number of monomers. By contrast, this contribution is always positive for both cyclohexane \ce{cC6H12} and n-hexane \ce{nC6H14}, larger for cyclohexane, and with no definite trend in terms of oligomer length. On the other hand, in all cases this contribution is one order of magnitude smaller than that of Coulomb interactions that is negative (that is attractive) for all three solvents. 
As a result, the energetic part of the non-covalent interactions is nearly identical for all three solvents and is attractive. The origin of this result is unclear. Likely, it can be ascribed to the hydrogen-oxygen interaction occurring between side chains upon $\pi-\pi$ stacking but it remains unclear how this contribution overwhelms the repulsive oxygen-oxygen contribution. What is clear, however, is that the energetic part of the interactions cannot be invoked to explain the differences in the folding processes of pPA in the three different folding. For this reason, we are going to tackle the calculation of the solvation free energy and its separation in energetic and entropic contribution in the next Section.
\subsection{pPA Solvation free energy }
\label{subsec:solvation_results}
\begin{figure}[htbp]
\centering
\begin{subfigure}[b]{0.45\textwidth}
    \centering
    \includegraphics[width=\textwidth]{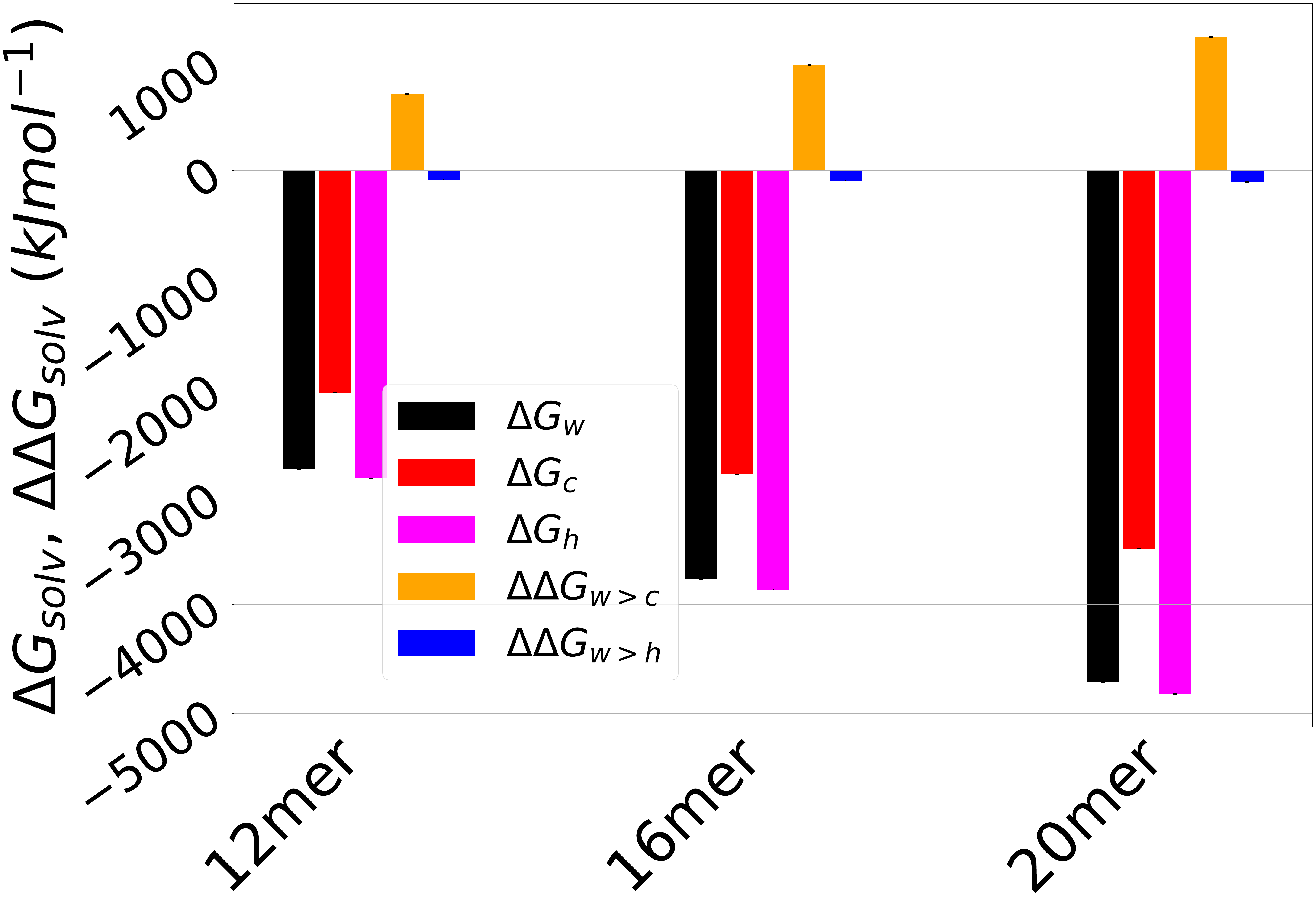}
     \caption{\label{fig:fig14a}}
    \end{subfigure}
    \begin{subfigure}[b]{0.45\textwidth}
    \centering
    \includegraphics[width=\textwidth]{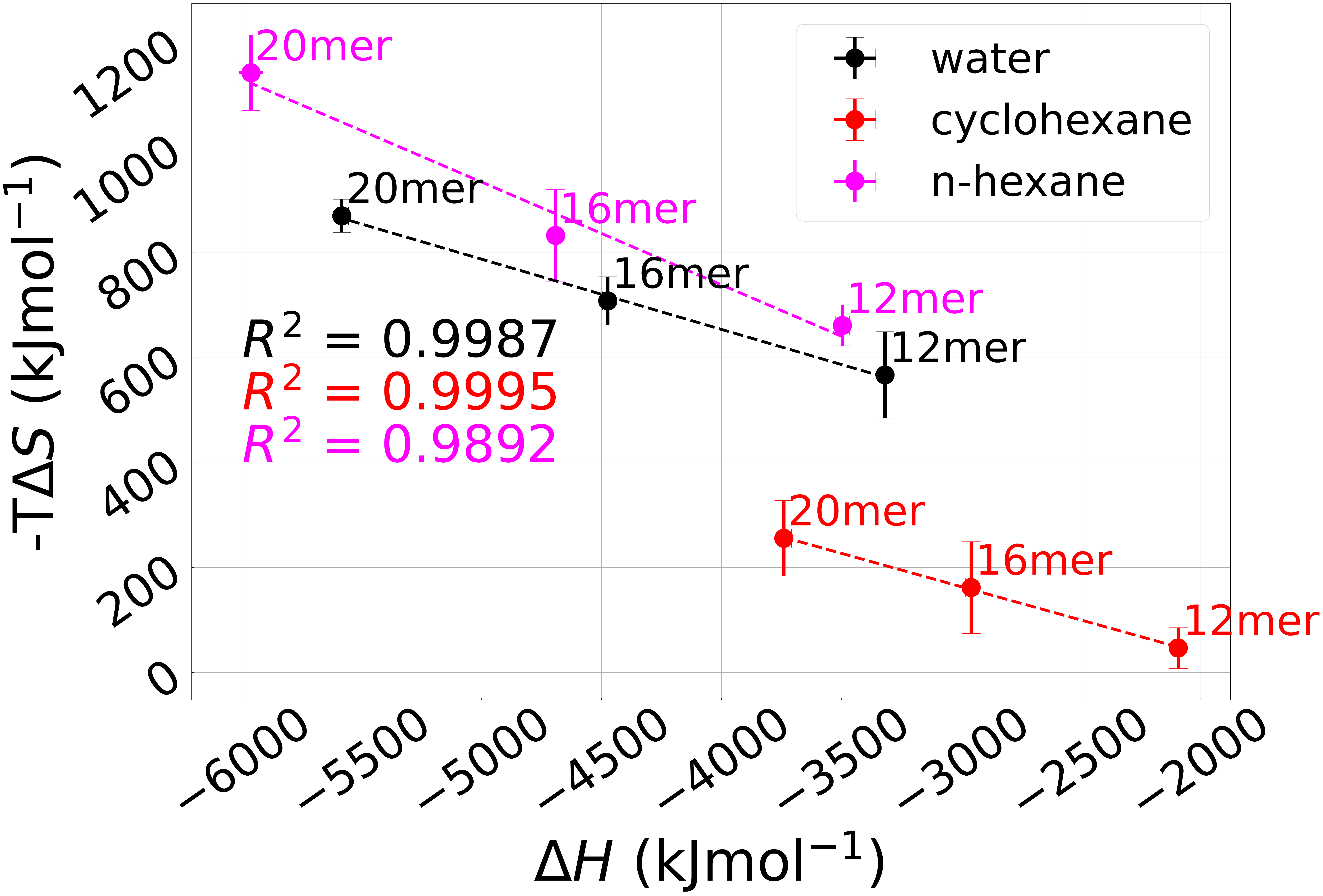}
     \caption{\label{fig:fig14b}}
    \end{subfigure}
  \caption{(a) Solvation free energy for pPA in the three different solvents studied in this work, water \ce{H2O} $\Delta G_{w}$ (black), cyclohexane \ce{cC6H12} $\Delta G_{c}$  (red), and n-hexane \ce{nC6H14} $\Delta G_{h}$  (magenta). Transfer free energies $\Delta\Delta G_{w>c} $ from water \ce{H2O} to cyclohexane \ce{cC6H12} (orange) and $\Delta\Delta G_{w>h}$ from water \ce{H2O} to n-hexane \ce{nC6H14} (blue) at $25^{\circ}$C are also reported. Each block refers to the considered polymer length : left \textit{12mer}, middle \textit{16mer} and right \textit{20mer}; (b) Entropic contribution $-T \Delta S$ of $\Delta G_{solv} $ as a function of the enthalpic counterpart $\Delta H$ in the case of  water \ce{H2O} (black) (\textbf{slope -0.133}), cyclohexane \ce{cC6H12} (red) (\textbf{slope -0.127}) and n-hexane \ce{nC6H14} (magenta) (\textbf{slope -0.195}).} 
  \label{fig:solv_contrib}
\end{figure}
The solvation free energy can be computed via thermodynamic integration, using Eq.(\ref{sec2:eq3}). Then $\Delta G_{\text{solv}}$ corresponds to the difference in free energy between pPA in a gas phase and in a specific solvent. We will denote as $\Delta G_{w}$, $\Delta G_{c}$, and $\Delta G_{h}$ the solvation free energies in water \ce{H2O}, cyclohexane \ce{cC6H12}, and n-hexane \ce{nC6H14}, respectively. The knowledge of these three solvation free energies allows also to compute $\Delta \Delta G_{w>c}$, and $ \Delta \Delta G_{w>h}$, that is the transfer free energy from water to either cyclohexane or n-hexane. A negative value of this quantity indicates a favorable solvation in either two solvents compared to water, the opposite for a positive value. As it also happens in the case of polypeptides \cite{Dongmo2023}, this relative quantity turns out to be the most relevant one, as it will be further discussed below.

Figure \ref{fig:fig14a} displays these free and transfer free energies, again for different oligomer lengths. While $\Delta G_{w}$, $\Delta G_{c}$, and $\Delta G_{h}$ are found to be all negative, indicating a stabilizing effect of all three solvents compared with the case of pPA in the gas phase, the relative transfer free energies turn out to be different. $\Delta \Delta G_{w>c}$ is positive and increasing with the oligomer length, whereas $\Delta \Delta G_{w>h}$ is negative, but significantly smaller in magnitude. This is a bit surprising in view of the putative predominant hydrophobicity of the pPA chain that would suggest a negative transfer free energy for both organic solvents. As the the chemical structure of the two organic solvent is nearly the same, the only alternative possibility is that the difference originates from the different shapes of the \ce{cC6H12} and \ce{nC6H14} molecules (see Figure \ref{fig:characteristic_lenght}), which in turn might affect the entropic contribution to the solvation free energy. To address this possibility, we have studied the temperature dependence of the solvation free energy from which it is possible to disentangle the energetic and the entropic contributions by first computing the entropy as the derivative with respect to temperature of the free energy, and then obtaining the enthalpy as a difference (see for instance Ref. \cite{Dongmo2023} for details).
This is done in Figure \ref{fig:fig14b} for all three solvents. The different behavior of cyclohexane \ce{cC6H12} and n-hexane \ce{nC6H14} is now clearly visible. We first note that in all three cases the slopes are negative and small. A zero slope would corresponds to the case of a process completely enthalpic dominated, a large slope to a process entropically dominated, a $-1$ slope to an optimal entropy-enthalpy compensation, that is what is actually occurring for polypeptides in water \cite{Dongmo2023} (see also next Subsection). The results of Figure \ref{fig:fig14b} then show that the three solvation processes are all mainly enthalpically driven, but the actual energies involved are rather different, with a large enthalpy gain for both water \ce{H2O} and n-hexane \ce{nC6H14} compared to cyclohexane \ce{cC6H12}. In the former case there is an entropy loss corresponding to the enthalpy gain, but with a much smaller magnitude. This confirms the enthalpic nature of the solvation process for all three solvents, and at the same time also confirms that n-hexane \ce{nC6H14} behaves differently from cyclohexane \ce{cC6H12} as a solvent for pPA and, intringuingly, more similarly to water \ce{H2O}. 

Once more, this behavior appears to be very different when contrasted with the solvation behavior of polypeptides in the same solvents \cite{Dongmo2023} in particular with one that was not considered in previous study \cite{Dongmo2023}, and that will be discussed in the next Section.
\subsection{Solvation free energy of polyphelylanaline (polyPHE) oligopeptides}
\label{subsec:polyPHE}

\begin{figure}[htbp]
\centering
    \includegraphics[width=0.5\textwidth, trim=3cm 26.5cm 3cm 27cm, clip=true, angle=0]{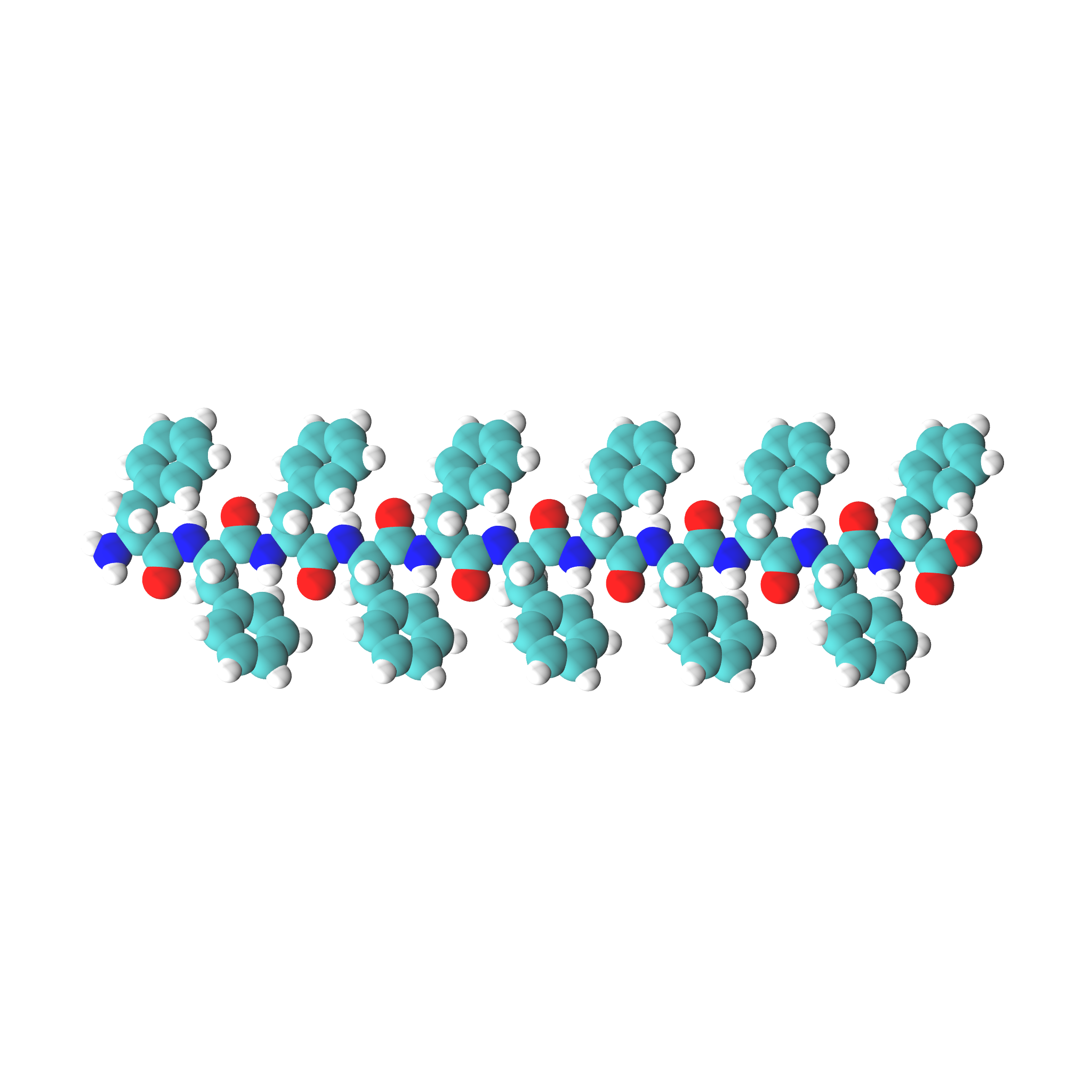}
  \caption{Initial stretched \textit{undeca}-phenylalanine (PHE) structure.} 
  \label{fig:phe11_strcut}
\end{figure}
\begin{figure}[htbp]
\centering
\begin{subfigure}[b]{0.48\textwidth}
    \centering
    \includegraphics[width=\textwidth]{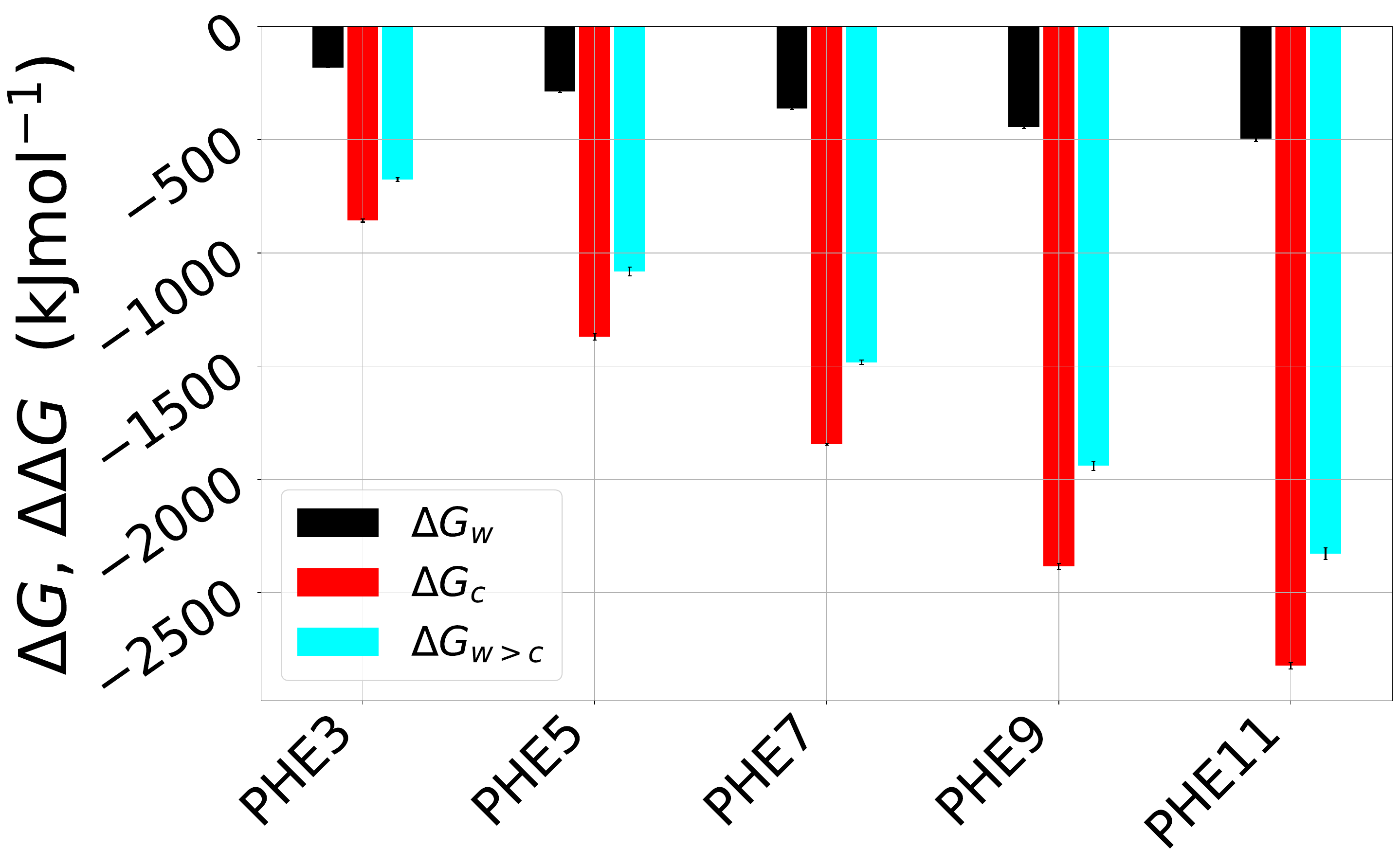}
     \caption{\label{fig:fig16a}}
    \end{subfigure}
    \begin{subfigure}[b]{0.48\textwidth}
    \centering
    \includegraphics[width=\textwidth]{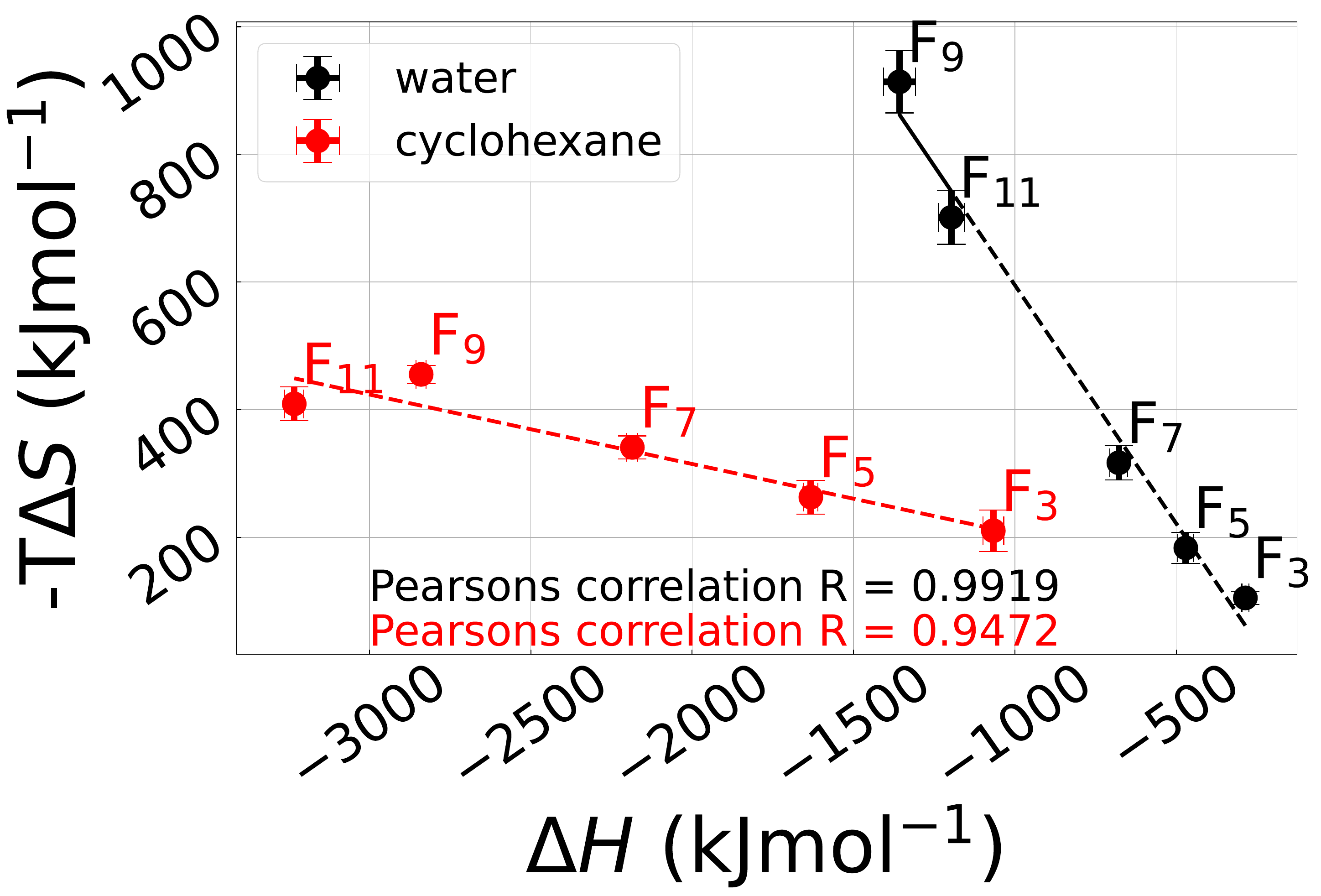}
     \caption{\label{fig:fig16b}}
    \end{subfigure}
  \caption{(a) Solvation $\Delta G_{s}$ and transfer $\Delta\Delta G_{s}$ free energy of polyphenylanaline oligopeptides of different length in water \ce{H2O} and in cyclohexane \ce{cC6H12} at $25^{\circ}$C. (b) Entropic contribution $-T \Delta S$ of the solvation free energy $\Delta G_{s} $ as a function of the enthalpic counterpart $\Delta H$ in the case of  water \ce{H2O} (black) (\textbf{slope -0.747}) and cyclohexane \ce{cC6H12} (red) (\textbf{slope -0.109}). Other relevant data are reported in Supplementary Material.}  
  \label{fig:polyF}
\end{figure}

An important outcome of our previous findings is that hydrogen bonds do not play any role in the stabilization of the final helical structure. This is at variance with polypeptides where intra-chain hydrogen bonding is known to be a key factor in the formation of the secondary structures \cite{Pauling1951}. Then it comes as no surprise that the helix formed by pPA has a morphology very different from those formed by biopolymers \cite{Banavar2023}. In our previous study \cite{Dongmo2023}, the solvation process of seven polypeptides was studied in both water and cyclohexane. Similarly to the present case, the solute polarity was found to control the folding, with the solvent polarity playing a more marginal role. Unlike the present case, however, the entropic contribution in water was found to play a pivotal role.

To clarify this point, we consider here the folding process of polyphenylalanine (PHE), one of the additional polypeptides that was not included in Ref. \cite{Dongmo2023}. One important feature of this peptide is to have a hydrophobic side chain due to its benzyl chemical structure, as illustrated in Figure \ref{fig:phe11_strcut}, whereas the backbone is significantly polar \cite{Dongmo2023}. Hence it is akin to an amphiphilic polymer \cite{Vasilevskaya2003} and it shares some similarities with pPA, albeit the distribution of the polar and hydrophobic moieties is very different from one another. Figure \ref{fig:fig16a} displays the solvation free energy in water \ce{H2O} $\Delta G_{w}$ of PHE with different number of residues from $3$ to $11$. In all cases it is large and negative. The same calculation in cyclohexane \ce{cC6H12} also reports a negative solvation free energy $\Delta G_{c}$ (Figure \ref{fig:fig16a}) even larger in magnitude. As a result the transfer free energy from water to cyclohexane $\Delta G_{w>c}$ is also negative (Figure \ref{fig:fig16a}) indicating that cyclohexane is a "poorer" solvent for PHE compared to water, and hence the corresponding folded state is more stable by a large margin. It is interesting to compare this result with the other hydrophobic polypeptides analyzed in Ref. \cite{Dongmo2023} (Glycine, Alanine, Isoleucine) that also display negative $\Delta G_{w>c}$ (see Figure 6 in Ref. \cite{Dongmo2023}) but nearly two orders of magnitude smaller, at the same temperature. The single enthalpic and entropic contributions reported in Figure \ref{fig:fig16b} (black for water and red for cyclohexane) also display a significant difference with each other and with those of the other studied hydrophobic peptides (see Figure 7 in Ref. \cite{Dongmo2023}). In water (black solid line) the enthalpic gain is slightly larger that the entropic loss, thus resulting into a small solvation free energy gain reported in Figure \ref{fig:fig16a} -- a value $\Delta G_{w}=0$ indicating an optimal enthalpy-entropy compensation \cite{Dongmo2023}. In cyclohexane (red line), the enthalpic gain is nearly equivalent to the solvation free energy gain and the entropic loss is significantly smaller, thus originating the large solvation free energy gain reported in Figure \ref{fig:fig16b}. A comparison with the results of other hydrophobic peptides (see Figure 7 in Ref. \cite{Dongmo2023}) also proves instructive. At variance with PHE, in Glycine, Alanine and Isoleucine the slopes of the two curves are nearly identical, albeit in water the values are much smaller than in cyclohexane, similarly to what we find here for polyphenylanaline PHE.
\section{Conclusions}
\label{sec:conclusions}
In this study we have addressed the relation between solvent \textit{quality}, solvent \textit{polarity}, and solvation free energy in the framework of the poly-polyphenylacetylene (pPA) polymer that has attracted considerable attention in the past in view of its important technological applications \cite{Nelson1997,Prince1999,Gin1999,Prince2000,Yang2000,Elmer2001,Sen2002,Elmer2004}. 
The equilibrium properties of pPA are of considerable interest because this foldamer  tend to adopt a helical structure in water and other solvents and this opens up interesting perspectives associated with the helicity \cite{Freire2016}. While this polymer can be classified as mainly hydrophobic, it forms hydrogen bond with water due to the presence of the polar moiety in its repeated unit. In order to be able to fully exploit these potential applications, however, a full control of the folding propensity is required. In this study, we specifically addressed this problem with a particular emphasis on the dependence on the solvent polarity, using the common practice of assuming the dielectric constant as a proxy for the solvent polarity \cite{Griffiths1979}. This study was carried out in several successive steps.

First, we have revisited the simple problem of a bead-spring Kramer-Grest polymer in explicit solvent \cite{Huang2021}, and shown that lowering the solvent quality is \textit{not} equivalent to lowering the temperature, as often tacitly assumed based on implicit solvent models. This simple example allowed us to identify the correct approach required to tackle the equilibrium properties of pPA in solvents with different qualities and polarities.

Following Ref. \cite{Smith1993}, we then assessed the 'hydrophobicity' of two paradigmatic organic solvents, cyclohexane \ce{cC6H12} and n-hexane \ce{nC6H14}, in terms of their mean force potentials in water. We found that the two solvents have similar hydrophobicities although with some quantitative differences related to their different shapes.


We then performed constant temperature and pressure molecular dynamics of (pPA)$_{m}$ in water \ce{H2O} (polar), in cyclohexane \ce{cC6H12}, and in n-hexane \ce{nC6H14}  for different number of monomers $m=12,16,20$.  Although there exist few previous experimental \cite{Nelson1997,Prince1999} and computational \cite{Elmer2001,Sen2002,Elmer2004} studies for similar polymers and solvents, to the best of our knowledge this is the first systematic computational study of this type.  We found that pPA forms stable helices in water (for sufficiently long oligomers), marginally stable in n-hexane, and unstable helices in cyclohexane.  
 
 As cyclohexane and n-hexane have similar chemical composition and  hydrophobicity (Figure \ref{fig:pmf_hydrophobicity}) but different shapes, their entropy changes upon folding of the polymer might be different \cite{Asakura1958}. The significant difference between water and the two organic solvents, and the small -- but relevant, difference between cyclohexane and n-hexane, was further assessed by measuring the distribution of the bending and dihedral angles that were found to reflect the above differences. Other order parameters such as the radius of gyration $R_g$, the root-mean-square-deviation (RMSD) from the initial extended conformation, and the solvent accessible surface area (SASA), confirm these findings. Using two of them ($R_g$ and RMSD) as "reaction coordinates", we further performed a free energy landscape analysis which provided consistency to this scenario with the existence of well defined deep minimum in the case of water, two broader and shallow minima in the case of cyclohexane, and something in between for n-hexane (Figure \ref{fig:fel}).

We further registered that the number of hydrogen bonds formed by pPA  with water is essentially unchanged during the folding process, with no formation of intra-chain hydrogen bonding. Hence, hydrogen bonds do not contribute to the stability of the folded state at variance of the case of peptides \cite{Sen2002,Dongmo2023}. The remaining non-covalent interactions (Lennard-Jones and Coulomb) were also monitored during the evolution. We find Lennard-Jones contribution to be negative in water, and positive in both cyclohexane and n-hexane. However, this contribution is largely overwhelmed by the Coulomb contribution that is negative for all three solvents, and significantly larger in magnitude. This suggests that the folding process is mainly enthalpically driven, with the main contribution electrostatic in nature, and largely independent of the solvent. As the chain folds driven by  $\pi-\pi$ stacking, the highly negatively charged oxygen atoms of each monomer likely combine with the positively charged hydrogen atoms of the adjacent turns, dictating the specific shape of the folded state.

Our study covers up $\approx$ 100 ns of the complete folding pathways from a linear random swollen coil, to a structured, ordered helical-like fold, in explicit solvents of different polarities, well beyond a previous computational study which could only assess the stability of the pre-organized oligomer fold, using implicit solvent and a different side chain \cite{Srikanta2002}. 

In the final step, we used thermodynamic integration to evaluate the solvation free energy of transfering a pPA single polymer from a gas phase to each of the three considered solvents. We found a negative value in all three cases, indicating a stabilizing effect of all three solvents. However, their relative stability was found to be different as assessed by the relative transfer free energies from water to cyclohexane (positive) and from water to n-hexane (negative).
By monitoring the temperature dependence of these solvation free energies it was possible to separate out the enthalpic and entropic contributions. In all three solvents, we found a large dominance of the enthalpic part, but with water and n-hexane behaving similarly and differently from cyclohexane, coherently with previous results.

The same calculation carried out on phenylalanile (PHE) oligopeptides underscores the significant differences with pPA. In this case, the PHE peptide shows the expected "entropy-enthalpy compensation" in water, with a nearly perfect anticorrelation dependence (slope $\approx -1$) in the $-T \Delta S$-$\Delta H$ plane. By contrast, in cyclohexane the peptide behaves similarly to pPA.

\textcolor{black}{The apparent insensitivity of the entropy/enthalpy ratio for pPA in all three solvents is quite puzzling, especially when compared with the response of oligopeptides such as the PHE analyzed here. It could be very well possible that pPA has a different response at temperatures outside the range considered in the present study (270-330 K) that is the typical range of interest for oligopeptides but not necessarily for synthetic polymers. These considerations certainly warrant future studies along these lines.  }

While not fully conclusive, our findings recapitulate and rationalize some open issues in past results on this class of synthetic foldamers. There exist other systems where similar features might be envisaged. For instance, it is known that the degree of association in Grignard compounds is strongly dependent on the details of the solvent \cite{Walker1969,DeGiovanetti2023}. We hope that our findings will trigger further efforts in all these systems that are far from being fully understood.

\textcolor{black}{All the reported simulation findings are from single-molecule runs. Motivated by present results, it would be very interesting to study the self-assembly properties of many pPA oligomers in the three different solvents considered in the present study. This is indeed part of the on-going effort by our group on the self-assembly of many chain polymer in solution, part of which has been already reported elsewhere \cite{Marcato2023,Arcangeli2024,Marcato2024}. We hope to be able to provide further insights on this system too in the near future.}
\clearpage
\providecommand*{\mcitethebibliography}{\thebibliography}
\csname @ifundefined\endcsname{endmcitethebibliography}
{\let\endmcitethebibliography\endthebibliography}{}

\end{document}